\documentclass[twocolumn]{aastex701}
\usepackage{xcolor}
\usepackage[T1]{fontenc}
\usepackage{amsmath}
\newcommand{\sectref}[1]{Section\,\ref{#1}}
\newcommand{\figref}[1]{Figure\,\ref{#1}}
\newcommand{\tabref}[1]{Table\,\ref{#1}}

\begin{document}

\title{Formation and eruption of a vortex-driven magnetic flux rope in the simulated quiet Sun}

\author[orcid=0000-0001-6438-7986]{Zhizhen Ye}
\email{502024260017@smail.nju.edu.cn} 
\affiliation{School of Astronomy and Space Science, Nanjing University, Nanjing 210023, China}
\affiliation{Key Laboratory of Modern Astronomy and Astrophysics (Nanjing University), Ministry of Education, Nanjing 210023, China}

\phantomsection\correspondingauthor{Feng Chen}
\email{chenfeng@nju.edu.cn}
\author[orcid=0000-0002-1963-5319]{Feng Chen}
\email{chenfeng@nju.edu.cn} 
\affiliation{School of Astronomy and Space Science, Nanjing University, Nanjing 210023, China}
\affiliation{Key Laboratory of Modern Astronomy and Astrophysics (Nanjing University), Ministry of Education, Nanjing 210023, China}

\begin{abstract}
Magnetic flux ropes (MFRs) are key structures for understanding flares and coronal mass ejections in active regions, but their characteristics in the quiet Sun remain poorly understood. With a radiative MHD simulation spanning from the upper convection zone to the corona, we analyze the formation and eruption of a supergranular-scale flux rope. We use a clustering method to group the closed field lines by their connectivity. The clusters are gathered further into 8 persistent cluster assemblies (CAs) that correspond to a flux rope and ambient magnetic structures. A persistent counterclockwise vortex is maintained at the converging point of the supergranules, i.e. the positive footpoint of the flux rope. The vortex continuously injects helicity into a magnetic flux tube and wraps ambient magnetic structures around the central flux tube. On a time scale of one hour, the flux tube evolves to a strongly twisted flux rope and erupts. The eruption exhibits similar current sheet structures as inferred from the flux rope eruption in active regions; however, the mass ejection and heating are much weaker and give rise to very insignificant observable features. The synthetic EUV images show mostly dimming features caused by density rarefaction in the expanding flux rope. This work helps elucidate the dynamic and complex evolution of flux ropes formed in the quiet Sun and suggests that similar events in the real Sun may have been underestimated due to their stealth behavior.

\end{abstract}

\section{Introduction}
Magnetic flux ropes are strongly twisted flux tubes that store excessive magnetic energy and are believed to be the magnetic structure driving solar eruptions, such as flares and coronal mass ejections (CMEs) \citep{Priest-2002A&ARv..10..313P,Aulanier-2012A&A...543A.110A,Janvier-2013A&A...555A..77J,Liu-2020RAA....20..165L,Cheng-2020ApJ...894...85C}. The crucial role of magnetic flux ropes in solar eruptions has led many previous studies on the formation of flux ropes in active regions or large-scale quiet Sun regions through flux emergence or convergence following photospheric shearing motions \citep{Leka-1996ApJ...462..547L,Fan-2001ApJ...554L.111F,Torok-2001sps..proc..364T,Ballegooijen-1989ApJ...343..971V,Klimchuk-1992ApJ...385..344K}. Such structures are expected to be more ubiquitous in the solar atmosphere. 

The solar photosphere hosts ubiquitous vortex motions on various scales, as discussed in the review by \citet{Tziotziou-2023SSRv..219....1T}. The magnetic field channels the vortex to the chromosphere and above, giving rise to magnetic tornadoes \citep{Wedemeyer-2012Natur.486..505W}, small-scale tornadoes in the quiet Sun \citep{Tziotziou-2018A&A...618A..51T}, and chromospheric swirls or vortices \citep{Wedemeyer-2012Natur.486..505W,Murabito-2020A&A...639A..59M}.

The quiet Sun is actually highly dynamic, where photospheric flow motions can continuously reshape the magnetic structure \citep{Balmaceda-2010A&A...513L...6B,Requerey-2017ApJS..229...14R,Requerey-2018A&A...610A..84R}. Photospheric vortex motions can drive rotating magnetic networks \citep{Zhang-2011ApJ...741L...7Z}, where flux ropes are created \citep{Yang-2015ApJ...803...86Y,Dai-2021ApJ...906...66D}. Mini-sigmoids, which are considered an observational counterpart of small flux ropes, were reported by \citet{Huang-2019ApJ...887..130H}. On a larger scale, rotating sunspots are often observed  \citep[e.g., ][]{Brown-2003SoPh..216...79B,Vemareddy-2012ApJ...761...60V,Ruan-2014ApJ...784..165R,Zhang-2022ApJ...933..200Z,Gou-2024SoPh..299...99G}.

The coupled evolution of the plasma and magnetic field through a vortex/magnetic flux tube through the difference layers of the solar atmosphere can be investigated in detail with magnetohydrodynamic simulations. Previous studies have analyzed the formation of flux ropes in the quiet Sun \citep{Amari-2015Natur.522..188A,Amari-2025ApJ...984L..37A,Robinson-2022A&A...668A.177R,Robinson-2023A&A...673A..79R,ChenYJ-2021A&A...656L...7C}, magnetic tornadoes and chromospheric swirls \citep{Kuniyoshi-2023ApJ...949....8K,Kuniyoshi-2024ApJ...969L..34K,Kuniyoshi-2025ApJ...990L..71K} as well as vortex tubes \citep{Skirvin-2025ApJ...988...18S}, Moreover, radiative MHD simulations, which account for sophisticated radiative and conductive transport, can provide plasma properties more consistent with those of the real Sun. This leads to benefits, such as synthetic observables that can be quantitatively compared with observations, and more realistic interactions between the plasma and magnetic field. With such simulations, \citet{Silva-2020ApJ...898..137S,Silva-2021ApJ...915...24S,Silva-2024ApJ...975..118S} investigated the interplay between vortices of the flow and magnetic fields, revealing the dependency on the magnetic-to-kinetic energy ratio.\citet{Battaglia-2021A&A...649A.121B} analyzed the correlation of chromospheric swirls and torsional Alfv\`en waves and quantified the upward energy flux. Some recent works \citet{Moreno-2025ApJ...995..137M} have demonstrated the formation of magnetic flux ropes through magnetic flux cancellation on granular scales, though the role of vortex motion was not explicitly discussed. 

In this study, we analyze the full evolution of a flux rope, which spontaneously forms as a result of persistent driving by a photospheric vortex and erupts in a fashion similar to that of large-scale flux ropes during major solar eruptions. The data are from the large-scale quiet Sun simulation shown in \citet{Chen-2022ApJ...937...91C}. In complement to previous studies of swirls and vortex-driven flux ropes that focus mostly on small-scales, the large 196\,Mm$^2$ horizontal area and the vertical domain, which spans from 18\,Mm beneath the photosphere to about 113\,Mm in the corona, allow us to investigate the complete chain of coupling between the vortex flows and magnetic field through several layers across a large stratification.

In previous studies, the quashing factor $Q$ \citep{Titov-2002JGRA..107.1164T, Titov-2007ApJ...660..863T} and twist number, which measure the winding of magnetic field lines around an axis \citep{Berger-2006JPhA...39.8321B, Liu-2016ApJ...818..148L} are often used as auxiliary parameters to isolate a coherent structure of interest, such as a magnetic flux rope, from the ambient magnetic field. Several efficient programs or algorithms have been developed to calculate the $Q$ factor \citep{Scott-2017ApJ...848..117S,Tassev-2017ApJ...840...89T,ChenJun-2022ApJ...937...26Z,Yang-2024zndo..13831036Y}. 

In the quiet Sun case, the magnetic field evolves dynamically with a turbulent velocity field, creating an environment that is largely different from the scenario considered in idealized theoretical models  \citep{Titov-1999A&A...351..707T}. The $Q$ factor in our quiet Sun corona exhibits numerous discrete structures, which means that it is almost impossible to determine the boundaries of a coherent flux rope solely based on the contour lines or iso-surfaces of the $Q$ factor, as has been done in many previous studies. 

To address this challenge, we design a novel procedure based on the clustering method to isolate and trace individual magnetic structures in a complex quiet Sun magnetic field. Clustering is a fundamental unsupervised machine learning technique that identifies natural groupings in unlabeled data. In solar physics, it has been applied to analyze Stokes profiles \citep{Viticchie-2011A&A...530A..14V}, detect and track active regions \citep{ML_Chen-2025ApJS..280...70C,ML_Caballero-2013SoPh..283..691C}, classify Mg~II flare spectra \citep{Panos-2018ApJ...861...62P}, and automatically segment coronal holes and active regions from EUV images \citep{Barra-2008AdSpR..42..917B,Barra-2009A&A...505..361B,Verbeeck-2014A&A...561A..29V,Inceoglu-2022ApJ...930..118I}. 

The paper is organized as follows. In \sectref{Methods}, we describe the setup of the simulation and the clustering method to identify magnetic structures. A comprehensive analysis of the formation and eruption of a vortex-driven flux rope is presented in \sectref{Results}. Our summary and discussion are given in \sectref{Summary and Discussion}.

\section{Data and Methods}\label{Methods}
\subsection{Simulation Data}\label{Simulation_Data}

In this study, we analyze the data of the quiet Sun stage in the simulations presented by \citet{Chen-2022ApJ...937...91C}. The simulation is conducted with the MURaM code \citep{Rempel-2017}, which solves fully compressive MHD equations with radiative transfer in the lower solar atmosphere, as well as optically thin radiative loss and field-aligned thermal conduction in the transition region and corona. The simulation domain spans 196.608\,Mm in both horizontal directions and 131.072\,Mm in the vertical direction. The domain is resolved by a $1024\times1024\times2048$ mesh, which yields grid spacings of 192\,km and 64\,km in the horizontal and vertical directions, respectively. The lateral boundaries are periodic. The top boundary is open, and the vertical mass flux in the two ghost cells is reduced by factors of 0.5 and 0.25. The bottom boundary is set at approximately 18\,Mm beneath the photosphere, which allows the development of convective motions at scales similar to those of supergranulation.

A small-scale dynamo operating in the convection zone generates mixed-polarity magnetic fields that permeate the corona. Continuous braiding by the granule motions yields an upward propagating energy flux, which is converted to plasma heating in the corona by viscous and resistive dissipation. The quiet Sun simulation evolves for more than 12 solar hours and maintains a stable million K hot corona with intermittent small-scale activities.

In this study, we focus on the time period between $t_{\mathrm{QS}}$=1~h~32~m~43~s and $t_{\mathrm{QS}}$=2~h~45~m~6~s, which captures the self-consistent generation and eruption of a small-scale magnetic flux rope, as briefly illustrated in \citet{Chen-2022ApJ...937...91C}. We set $t_{\mathrm{QS}}$=1~h~32~m~43~s as the reference time $t_{0}$ in this study, all the time instants mentioned
hereafter are relative to $t_{0}$. The periodic horizontal domain has also been shifted so that the region of interest is located in the center of the domain.

\subsection{Clustering Method to Identify Magnetic Structures}
\label{subsection_cluster_method}
Many studies have revealed the importance of magnetic field line connectivity and topological attributes. Therefore, we introduce a new method to identify magnetic structures based on the locations of field line footpoints. The photospheric magnetic field in the quiet Sun can be divided into network and inter-network fields, which form positive/negative magnetic concentrations of various sizes on the photosphere. We infer that the connectivity—defined by the locations of the positive and negative footpoints of a field line—will also appear as clustered distribution in an abstract four‑dimensional parameter space. This makes a clustering method well suited. HDBSCAN (Hierarchical Density‑Based Spatial Clustering of Applications with Noise) is an advanced clustering algorithm that can find arbitrarily shaped, varying‑density clusters without requiring a predefined number of clusters or other prior information \citep{campello2013density, mcinnes2017hdbscan}, making it an ideal tool. 

The clustering method used in this study consists of four steps:
\begin{enumerate}
  \item Place 600,000 seed points on the horizontal slice at $z=0.304$\,Mm over the full simulation domain, with $|b_z|$ as weight. Then compute the magnetic field lines passing through these seed points. 
  \item Consider only the closed field lines from the previous step and record their positive footpoints \footnote{The footpoint in the positive polarity} $(x_p,y_p)$ and negative footpoints $(x_n,y_n)$ on the photosphere.
  \item Cluster the connectivity vector $(x_p, y_p, x_n, y_n)$ using the HDBSCAN algorithm, which yields more than 4000 clusters for the entire simulation domain.
  \item Merge clusters whose centroids lie within a constant range into cluster assemblies (CAs). Eight persistent CAs are selected to study the temporal evolution of the magnetic field.
\end{enumerate}

We set the parameter $minimal\_cluster\_size=25$ and adopt the Euclidean distance option. The unsigned magnetic flux on the photosphere over the whole domain is $2\times10^{22}$\,Mx, and the minimal cluster contains approximately $1\times10^{18}$\,Mx magnetic flux, which is comparable to the typical flux of an inter-network concentration. This parameter choice allows us to identify weak magnetic loops; however, it may split strong magnetic loops into several clusters. Therefore, we add step 4, which is a postclustering process, to recombine clusters with similar connectivities into cluster assemblies (CAs).
 
This procedure identifies the CAs of interest at each time step and does not assume any relation between the CAs at different times. However, compared with the coronal magnetic field, the photospheric footpoints of the field lines evolve much more slowly. During the time period considered here, the locations of the photospheric flux concentrations remain almost the same. Consequently, a group of field lines rooted in the same flux concentration effectively captures the same CAs throughout the evolution.

\section{Results}
\label{Results}
\subsection{Identified Persistent Cluster Assembles of Field Lines}
\begin{figure*}
  \centering
  \includegraphics[width=18cm]{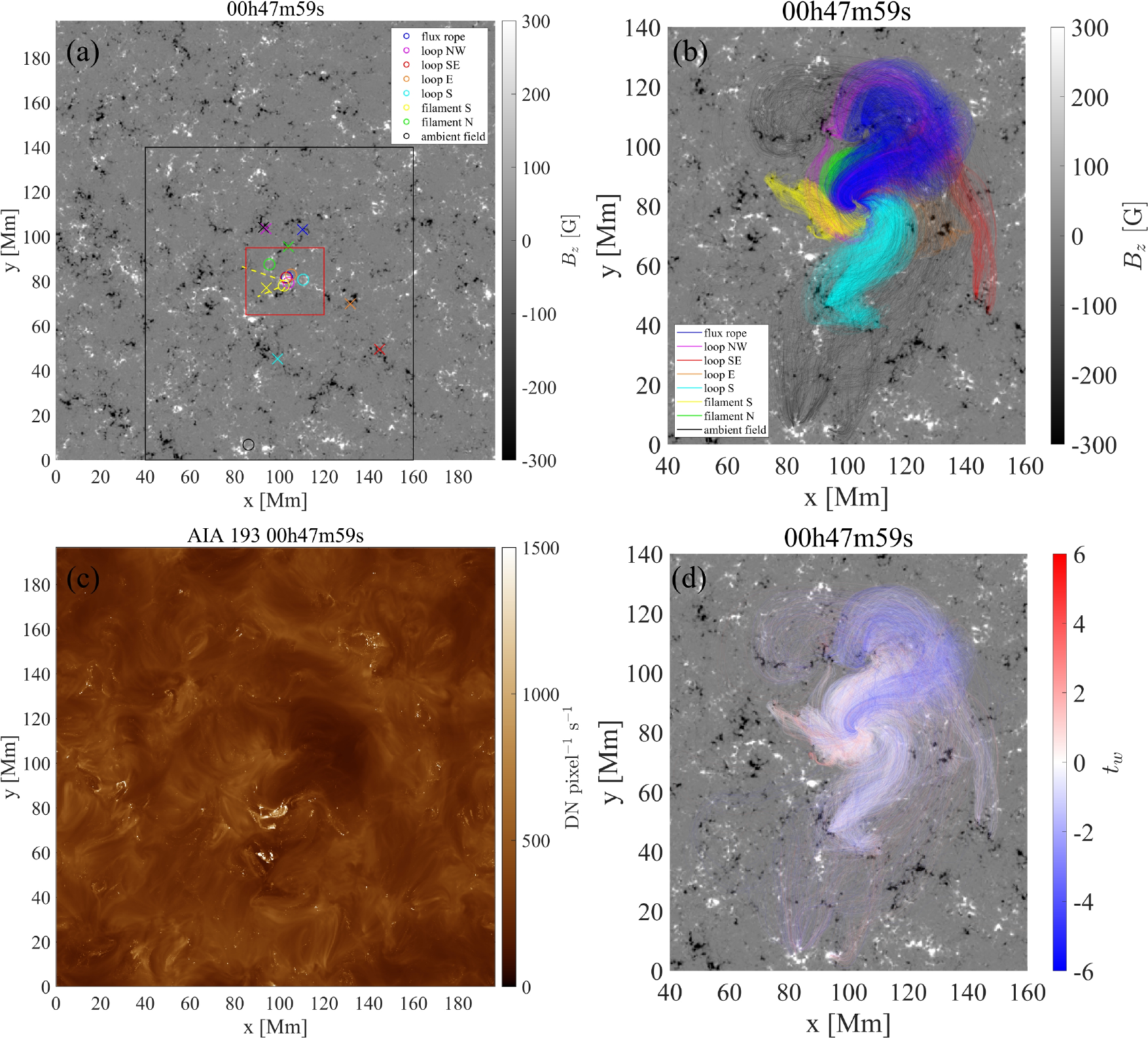}
  \caption{(a) Photospheric magnetogram of the full domain, the colored circles represent the centroids of positive/negative footpoints of different CAs. The yellow solid and dashed lines represent the two types of field lines in \textit{filament S}. The black rectangle contains the 8 identified CAs. The red rectangle includes the positive footpoints of all the CAs except \textit{ambient field}.
  (b) Top view of 8 CAs identified by the clustering method. The field-of-view corresponds to the black rectangle in Panel (a).
  (c) Synthetic EUV image of the full domain in the AIA 193 \AA~channel.
  (d) The same field lines in Panel (b) but colored by their twist number. An animation that covers the dynamics evolution of 72 minutes starting from $t=t_0$ is available.
 (An animation of this figure is available. )}
  \label{fig:aia_magneto_cluster_twist}
\end{figure*}

An overview of the magnetic structures identified following the clustering method is shown in \figref{fig:aia_magneto_cluster_twist}. We selected 8 persistent CAs relevant to the eruption region, including the flux rope, 4 nearby groups of coronal loops, 2 low-lying filaments, and the overlying large magnetic arch. The centroids of their footpoints are listed in \tabref{tab:all_CAs_infor} and shown in \figref{fig:aia_magneto_cluster_twist} (a). All names of the CAs are in italics throughout the remainder of the paper.

\begin{deluxetable}{lccr}
\tablecaption{Selected persistent CAs}
\label{tab:all_CAs_infor}
\tablehead{
    \colhead{Cluster assembles } & \colhead{Positive footpoint centroid  ($x_p$,$y_p$)} & \colhead{Negative footpoint centroid ($x_n$,$y_n$)} & \colhead{Designated Color}
}
\startdata
\textit{flux rope} & (103,81) & (109,103) & blue \\
\textit{loop NW}     & (105,79) & (96,100)  & pink \\
\textit{loop SE}     & (103,81) & (144,50)  & red \\
\textit{loop E}      & (104,82) & (134,70)  & orange \\
\textit{loop S}      & (111,81) & (99,47)   & cyan \\
\textit{filament S}  & (101,77) & (92,78)   & yellow \\
\textit{filament N}  & (96,88)  & (104,96)  & green \\
\textit{ambient field} & (86,7) & (91,104)  & black
\enddata
\end{deluxetable}

Panel (b) of \figref{fig:aia_magneto_cluster_twist} presents the top view of the region marked by the black rectangle in Panel (a), which encloses the magnetic field lines in the identified CAs. Panel (c) shows the synthetic AIA 193\,\AA~ image, in which a dimming region can be seen cospatially with \textit{flux rope}. The animation of \figref{fig:aia_magneto_cluster_twist} demonstrates that the dimming region expands simultaneously with \textit{flux rope}, and the details of synthetic emission are discussed in \sectref{subsect:Observable Features of the Flux Rope Eruption}. Panel (d) shows the same field lines as panel (b) but colored according to their twist number $t_{w}$, which is calculated by  
\begin{equation}
t_w=\int{\frac{(\nabla\times \vec{B})\cdot \vec{B}}{4\pi B^2}}{dl}.
\end{equation}
\textit{flux rope} is significantly twisted. The values of $|t_{w}|$ within this assembly are generally greater than 2 and can approach 6, which are significantly larger than those typically found in active region flux ropes.

Another notable point is that field lines may cross the photosphere many times and we removed the convective zone during the analysis so that the field lines crossing the photosphere are cut into segments. In Panel (a) of \figref{fig:aia_magneto_cluster_twist}, the yellow line and dashed line show the two types of field lines in \textit{filament S}: the straight line represents the field lines directly connecting the positive and negative concentrations at the sides, whereas the dashed line represents field lines connecting the small flux concentrations inside. This can be recognized as the combination of a magnetic loop and the underlying filament.
 
\subsection{Temporal Evolution of the Flux Rope}

\begin{figure*}
  \centering
  \includegraphics[width=14cm]{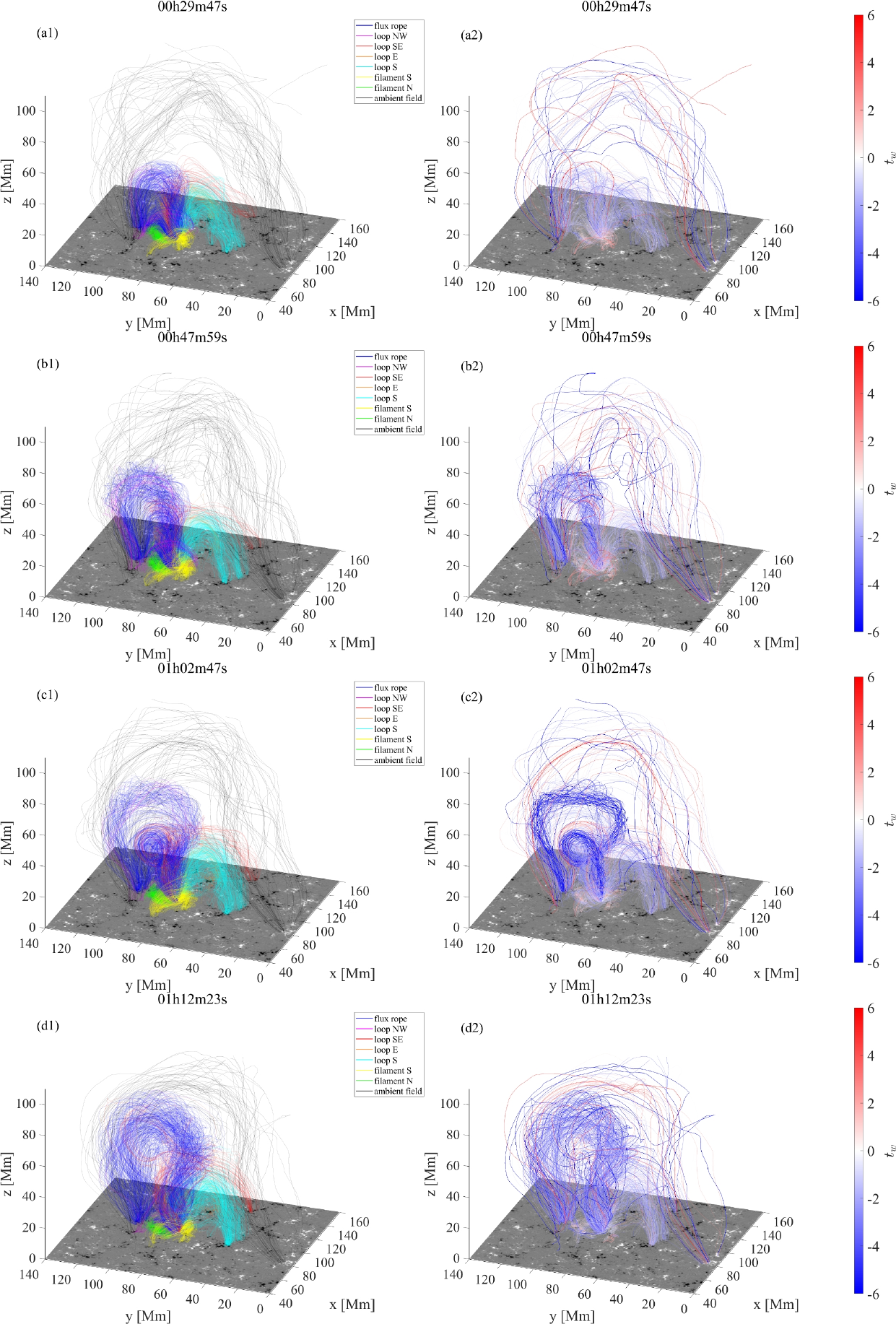}
  \caption{Evolution of eight CAs. Left column: different CAs represented by designated colors. Right column: same field lines but colored by the twist number. Rows top to bottom: (a) Initial normal coronal loop (b) Rotating flux rope accumulates twist. (c) Unstable flux rope forms a writhe shape. (d) Untwisting eruptive flux rope. An animation that covers the dynamics evolution of 72 minutes starting from $t=t_0$ is available.
  (An animation of this figure is available.)}
  \label{fig:twist_all}
\end{figure*}

\begin{figure*}
  \centering
  \includegraphics[width=16cm]{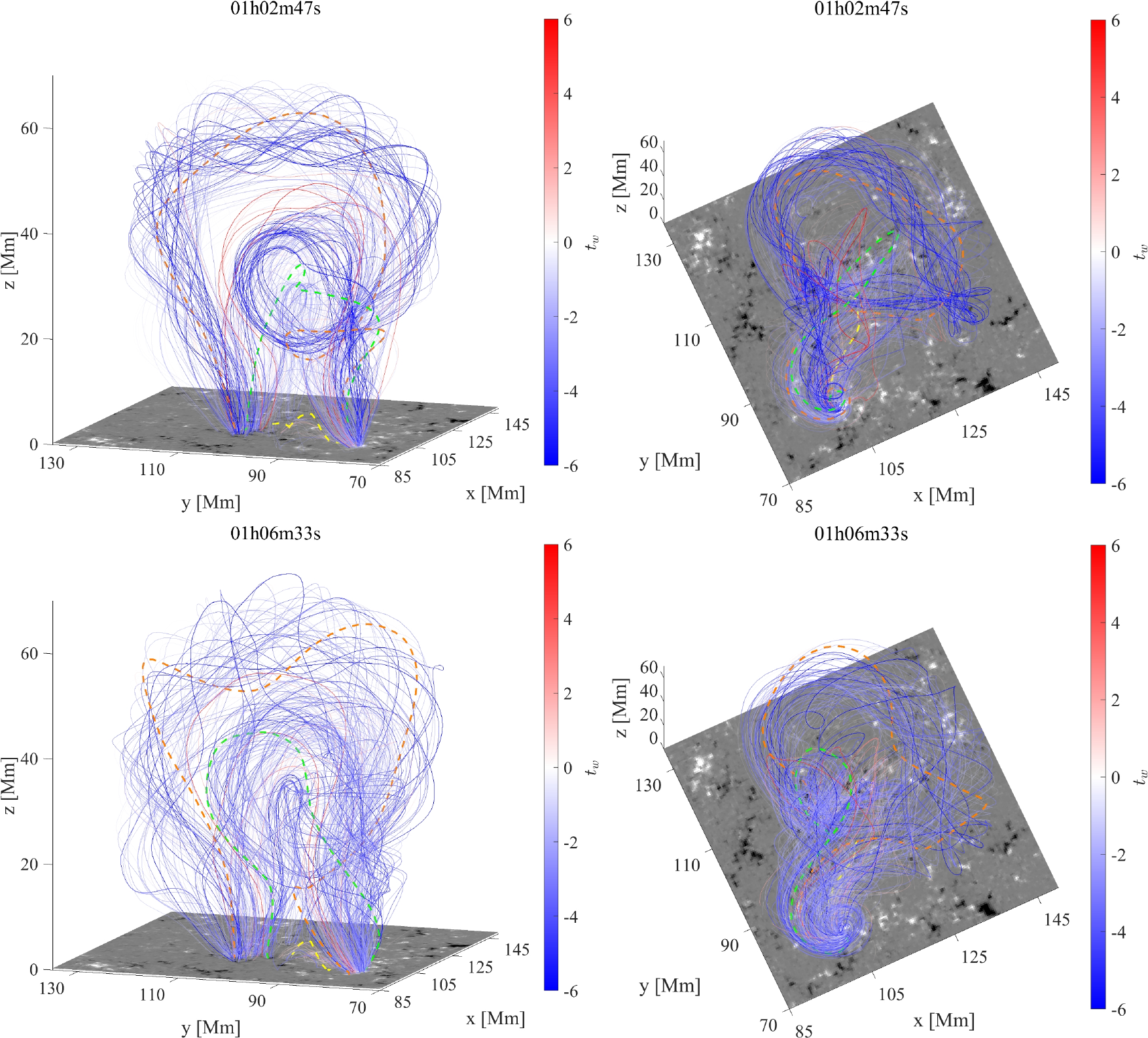}
  \caption{Two views of \textit{flux rope} before and after the internal reconnection. Field lines are colored by the twist number. The orange, green, and yellow dashed lines highlight field lines in three bundles (see main text for details) in this cluster assembly.} 
  \label{fig:twist_FR}
\end{figure*}

We examine how a magnetic flux bundle evolves from a weakly twisted tube to an erupting flux rope\footnote{We randomly select 10\% of field lines in each CA to gain better visualization effect in \figref{fig:twist_all} and \figref{fig:twist_FR}}. \figref{fig:twist_all} shows the magnetic field lines of all the CAs and their twist numbers at four representative time instants. The general magnetic configuration remains the same, \textit{Flux rope} lies beneath a closed envelop field, which is indicated by \textit{ambient field}, which is naturally formed as a result of the large-scale flux imbalance. The negative footpoint of \textit{ambient field} is close to the negative footpoint of \textit{flux rope}, which implies that the interactions between the two CAs may occur well before eruption onset. \textit{Loop NW} partly illustrates the interface between \textit{flux rope} and \textit{ambient field}. \textit{Flux rope}, \textit{loop SE}, and \textit{loop E} share nearby positive footpoints but connect to much more distant negative footpoints. Consequently, this pattern implies the presence of QSLs between each pair of CAs, extending from the photosphere into the corona. \textit{Filament S} and \textit{filament N} are low-lying filaments that host cool plasma. \textit{Loop S} is a magnetic loop with a relatively close positive footpoint but a negative footpoint opposite to that of \textit{flux rope}.

Panel (a) of \figref{fig:twist_all} shows that \textit{flux rope} is a normal magnetic loop that reaches 40\,Mm height when the dimming region begins to expand, as discussed in \sectref{subsect:Observable Features of the Flux Rope Eruption}. 
Panel (b) shows the twisting of \textit{flux rope} and surge of \textit{filament S} when the vortex flux reaches a maximum and twist number increases most rapidly.
Panel (c) shows that the most twisted \textit{flux rope} is in an unstable state, and a coherent bundle of field lines with extraordinarily high $t_{w}$ (generally greater than 4) can be clearly identified.
Panel (d) shows the eruptive \textit{flux rope} rapidly rising and pushing against the overlying \textit{ambient field}, and the decrease in the twist number is the result of magnetic reconnection. 

We can roughly define three evolutionary phases. In the first phase, the \textit{flux rope} interacts with \textit{loop SE} and \textit{loop E} and splits into upper and lower parts. The upper part not only twists slowly but also stretches along the +x direction and rises gradually along the +z direction, while the lower part remains relatively quiescent. In the second rapid-twisting phase, the location of the upper part remains approximately unchanged, but its twist increases dramatically, forming an unstable flux rope. Finally, reconnection is triggered.

We must emphasize that \textit{flux rope} does not evolve through a simple process of untwisting and eruption. Instead, complex internal reconnection takes place: a portion of the lower field‑line bundle subsequently gains twist from the most twisted field lines and becomes the main body of the eruptive flux rope. This evolution can be seen in the animation associated with \figref{fig:twist_all}. Notably, the main body of \textit{flux rope} appearing in panel (c) and that in panel (d) of \figref{fig:twist_all} are two distinct components, located at different three‑dimensional locations, their apparent resemblance is merely due to the viewing perspective.

In the last phase, \textit{flux rope} also has significantly expanded such that the field lines in the same CA may come close, which facilitates internal reconnections in this CA. \figref{fig:twist_FR} shows a zoomed-in view of two snapshots from two viewing angles just before and after internal reconnection, where only field lines within \textit{flux rope} are displayed. 
In the top row of \figref{fig:twist_FR}, \textit{flux rope} consists of three components. One is a strongly twisted field line bundle that reaches a height of 60\,Mm and extends in the +x direction, and the most twisted field lines appear as a writhed structure of approximately one turn. The other one is a bundle of weakly-twisted and lower field lines that connects the footpoints of \textit{flux rope} via a more direct path (in contrast to the former bundle). Finally, the underlying field lines along the line extend between the negative and positive polarity concentrations. The thick dashed lines with different colors in \figref{fig:twist_FR} highlight the axis line of three bundles and the occurrence of the reconnection. 
In the bottom row (4 minutes later) of \figref{fig:twist_FR}, the twist number along the upper and writhed bundle clearly decreases, whereas the twist number along the lower and direct-connection bundle significantly increases. A similar scenario of internal reconnection has been demonstrated in the model \citet{Aulanier-2012A&A...543A.110A}. The internal magnetic reconnection transfers the twist to the underlying bundle of magnetic field lines and drives its rapid eruption.

\subsection{Thermodynamic properties of the flux rope and current sheet}

\begin{figure*}
  \centering
  \includegraphics[width=18cm]{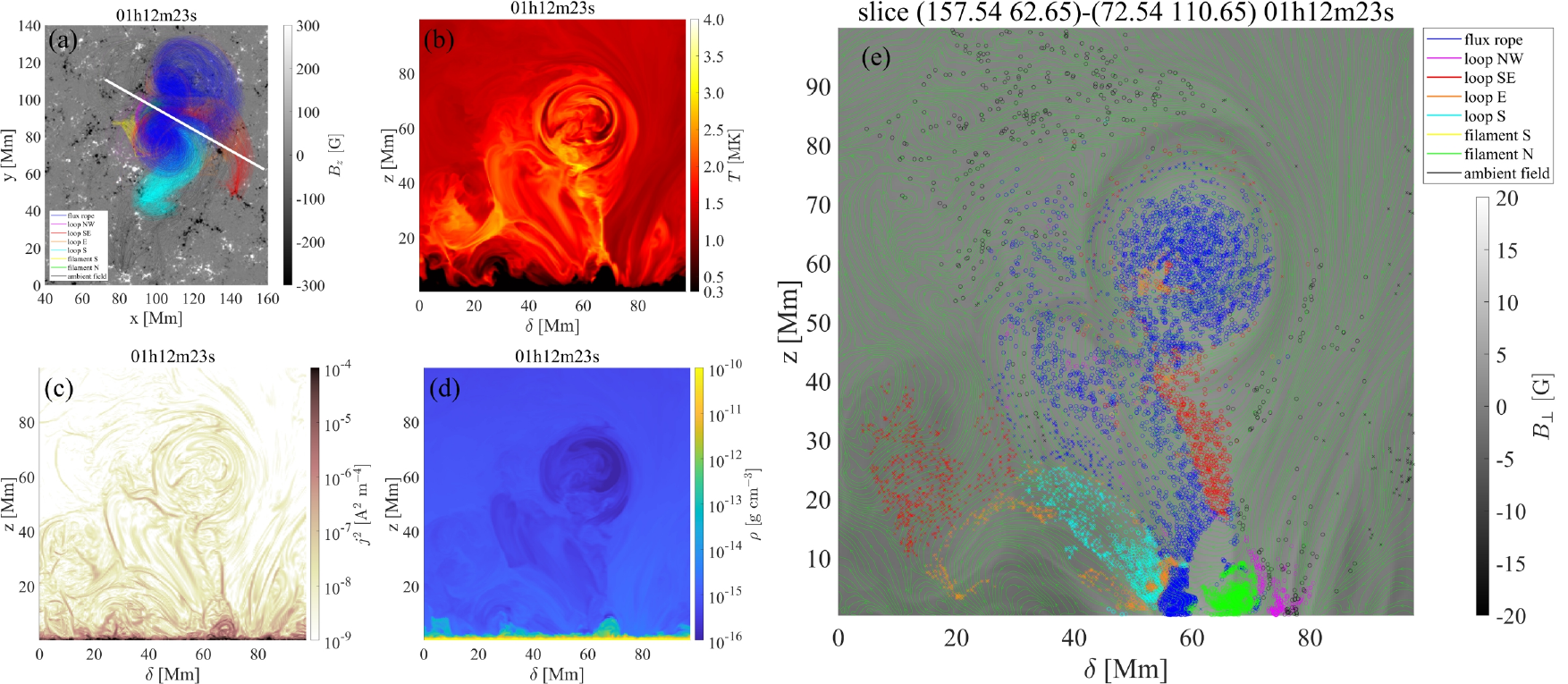}
  \caption{Temporal evolution of the cross section of \textit{flux rope}. Panel (a) shows the location of a static vertical slice. Panels (b), (c), and (d) show the distributions of temperature, squared current, and density on the slice, respectively. The horizontal axis of the slice ($\delta$) points from larger $x$ to smaller $x$. Panel (e) shows the intersections of different CAs with the slice; the gray background represents the magnetic field strength normal to the slice, and the green streamlines represent the magnetic field parallel to the slice. An animation that covers the dynamics evolution of 72 minutes starting from $t=t_0$ is available.
  (An animation of this figure is available.)}
  \label{fig:slice_PIL}
\end{figure*}

We place a vertical slice cutting approximately through the middle of \textit{flux rope}, as shown in Panel (a) of \figref{fig:slice_PIL}. The direction of the slice is chosen so that it is perpendicular to the axis of \textit{flux rope}. The width and height of the slice are denoted by $\delta$ and $z$, respectively. The origin and end points of the $\delta$ axis are $(x,y)=(157.54,62.65)$ and $(x,y)=(72.54,110.65)$, respectively.

Panels (b), (c), and (d) present the temperature ($T$), current density squared ($\mathbf{j}^{2}$), and plasma density ($\rho$) in the slice, respectively. Panel (e) shows the magnetic field component perpendicular to the slice. The image is overlaid with streamlines of the magnetic field projected onto the slice. The colored dots and crosses indicate the intersections of different CAs (obtained from the clustering method) with the slice. This panel illustrates that CAs originating from different photospheric flux concentrations form coherent structures that are interwoven in the corona.

The animation of \figref{fig:slice_PIL} presents the dynamic evolution of the plasma properties and the magnetic field during the eruption. From the dots and crosses in Panel (e), we can see that not only the circular magnetic field component of \textit{flux rope} but also the other CAs rotate together and rise. 

At the moment shown in \figref{fig:slice_PIL}, which is in the rapid eruption phase, We can clearly identify the main body of eruptive flux rope and the remaining part of field line bundle that stretched in the +x direction. A strong current sheet forms below the erupted \textit{flux rope} ($\delta \approx 75$) and connects to low-lying short loops. Despite its smaller spatial scale, the overall magnetic configuration appears very similar to those proposed in classical models of flares and CMEs. In addition, two slightly weaker current sheets ($\delta \approx 60$) can also be found in the vicinity of the strongest one, reflecting the more complex nature of the quiet Sun magnetic field, where multiple groups of field lines are involved in the eruption. A comparison with the magnetic field intersections in the slice shown in Panel (e) suggests that these current sheets are cospatial with the QSLs among \textit{flux rope}, \textit{loop SE}, and \textit{ambient field}.

We see many fine structures in the temperature map shown in Panel (b). Some cold photospheric plasma is confined to the flux rope and carried into the corona during eruption. Some thin threads, which are cospatial with the enhanced current density, reach temperatures above 3 MK, indicating efficient heating by magnetic energy dissipation. However, the mean temperature of \textit{flux rope} is approximately 2 MK, which is not significantly hotter than the ambient corona. Moreover, the main body of \textit{flux rope} clearly has a lower density than the ambient corona does, as shown in \figref{fig:slice_PIL}(d), which explains why it would be observed as propagating dimming. The heated hot threads also have rather low density and thus do not yield evident observable features in the synthesized EUV images.

\subsection{Photospheric Vortex and Flux Rope Formation}
\begin{figure*}
  \centering
  \includegraphics[width=18cm]{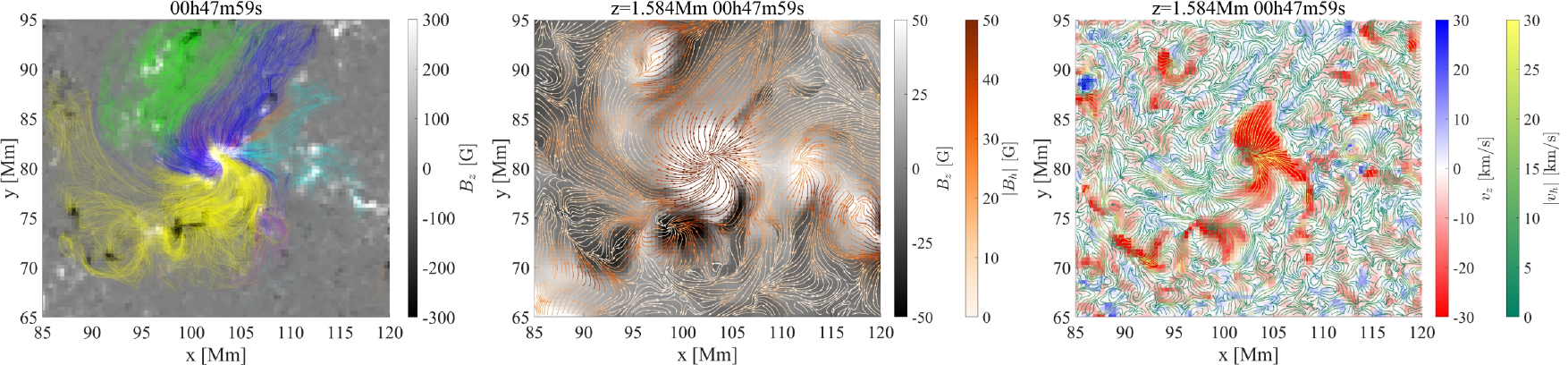}
  \caption{Left panel: top view of field lines below 3\,Mm for different CAs. The field of view is the same as the red rectangle in panel (a) of \figref{fig:aia_magneto_cluster_twist}. Middle panel: magnetic field at $z = 1.584$~Mm. The grayscale shows the strength of the vertical component of the magnetic field, and the orange streamlines show the direction and strength of the horizontal magnetic field. Right panel: velocity field at the same height, using the same representation for the flow field but with a different color map. An animation that covers the dynamics evolution of 72 minutes starting from $t=t_0$ is available.
  (An animation of this figure is available.)}
  \label{fig:swirl_show}
\end{figure*}

\begin{figure*}[p]
  \centering
  \includegraphics[width=16cm]{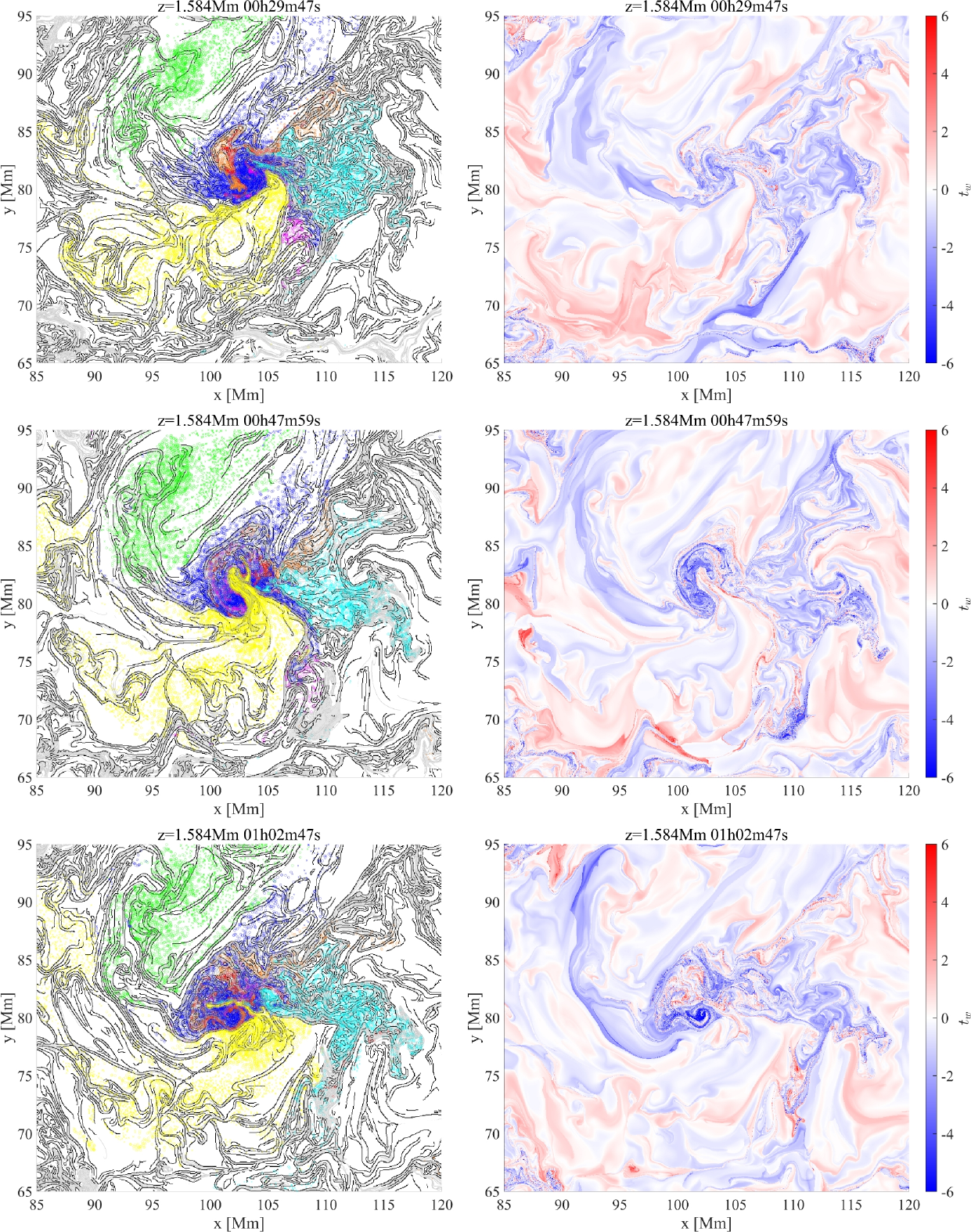}
  \caption{Comparison of our clustering method with the $Q$ factor and twist number at three time instants. Left column: the gray-filled contours  represent regions where $Q > 10^{4}$. The black lines represent the edges of the $Q$ value recognized by the Canny algorithm \citep{Canny-1986ITPAM...8..679C}. The colored markers represent the intersections of all the CAs with the horizontal slice at $z = 1.584$~Mm; the `o' and `x' markers indicate field lines crossing the slice upward or downward, respectively. Right column: twist number on the same slice. An animation that covers the dynamics evolution of 72 minutes starting from $t=t_0$ is available.
  (An animation of this figure is available.)}
  \label{fig:Q_cluster_tw}
\end{figure*}

\begin{figure*}
  \centering
  \includegraphics[width=18cm]{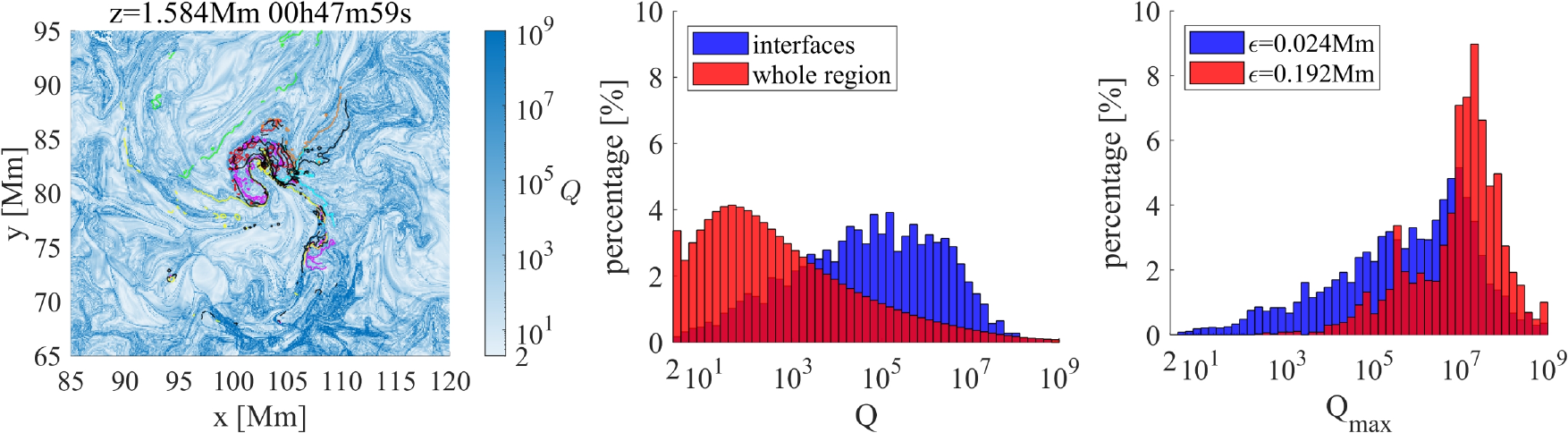}
  \caption{Comparison between QSL and CA interfaces. The left panel shows a map of $Q$ and CA interfaces in an extended area around the \textit{flux rope}. Colored lines represent the interface of the \textit{flux rope} and the CA with the designed color. The black lines represent the interfaces between other CAs except the \textit{flux rope},  The middle panel shows the histograms of $Q$ values at the CA interfaces and those over full domain in the left panel. The right panel shows histograms of the maximum $Q$ within an epsilon‑neighborhood of each pixel at the CA interface. Two $\epsilon$ values have been tested.}
  \label{fig:qsl_interface}
\end{figure*}

\begin{figure*}
  \centering
  \includegraphics[width=18cm]{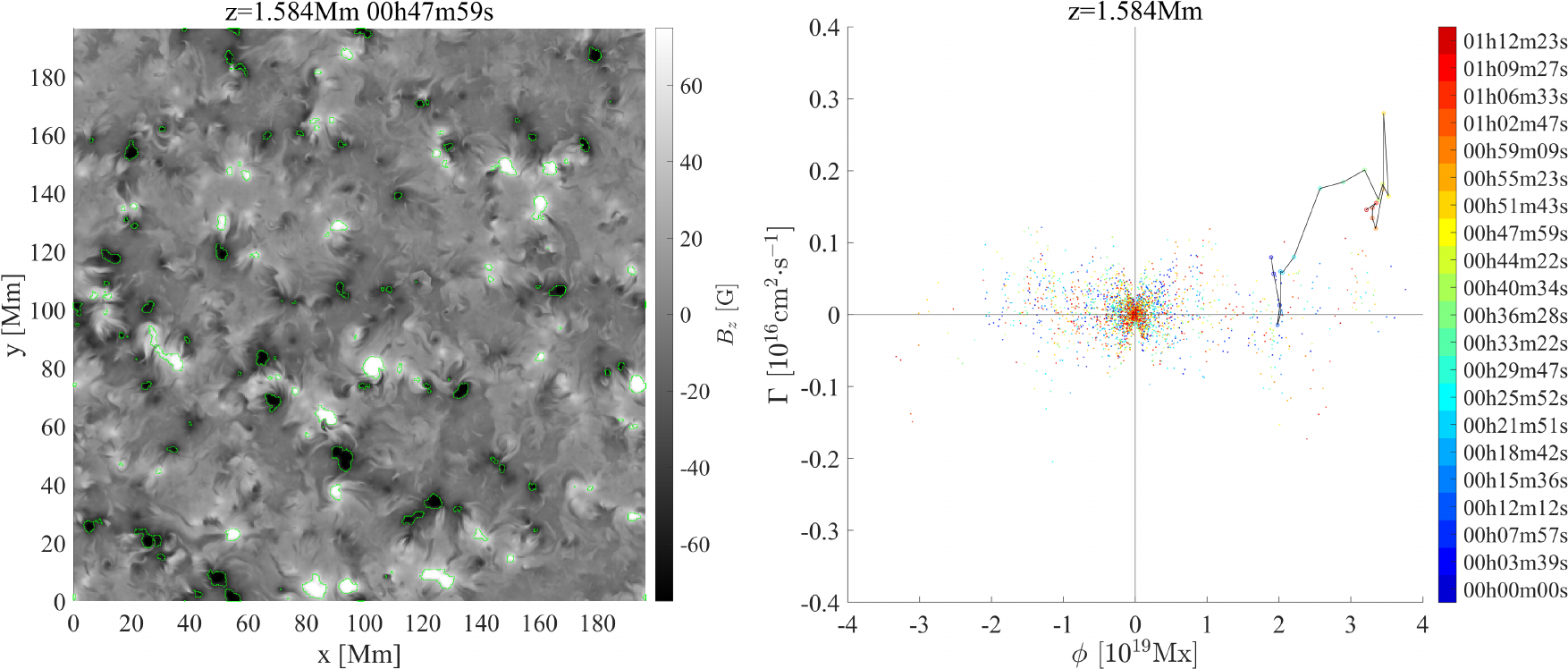}
  \caption{Left panel: spatial distribution of $\mathbf{|B_z| > 50}$~G concentrations at $z = 1.584$~Mm in the full simulation domain. The green contours mark the edges of all concentrations that evolved in the statistical analysis.} Right panel: Joint distribution of magnetic flux and vorticity flux for all concentrations at all time instants. Different colors of dots represent concentrations at different times, while the circles connected by the black line represent the temporal evolution of the core positive concentration.
  \label{fig:swirl_dist}
\end{figure*}

We present the crucial role of photospheric vortex motions in driving the formation and eruption of \textit{flux rope}. \figref{fig:swirl_show} shows the impact of vortex motion on the positive footpoint of \textit{flux rope}. The fields of view of the three panels are identical and correspond to the red rectangle in panel (a) of \figref{fig:aia_magneto_cluster_twist}. The left panel shows the photospheric magnetogram and the field lines below 3\,Mm belonging to all the CAs;  \textit{ambient field} does not enter this volume. The positive footpoints of five CAs (\textit{flux rope}, \textit{loop NW}, \textit{loop SE}, \textit{loop E}, and the majority of \textit{filament S}) are rooted in the same strong positive concentration located at $(x, y) = (103, 81)$. The positive footpoints of \textit{loop S} and \textit{filament N} are rooted in two other nearby concentrations.

The middle panel presents the magnetic field on a horizontal slice at $z = 1.584$\,Mm. The map of the vertical magnetic field strength is overlaid with colored streamlines that outline the horizontal magnetic field, indicating both direction and strength simultaneously. The right panel shows the velocity field at the same height but with a different color scale.

Flux concentrations in the quiet Sun are usually found cospatial with the downflow lanes of convective cells of various sizes. The same scenario occurs here: the downflow in the flux concentration induces a continuous drainage of plasma along the fanning‑out field lines. Moreover, the converging horizontal flows evolve into a persistent counterclockwise vortex, as clearly demonstrated by the swirling streamlines in the area centered at $(x, y) = (103, 81)$. We can also distinguish the vortex motion by the counterclockwise rotation of the intruding negative‑polarity concentration.

This continuous vortex motion drives the formation and eruption of the highly twisted \textit{flux rope}. In \figref{fig:Q_cluster_tw}, we present the evolution of the comparison of the $Q$ factor (left column), the magnetic twist number (right column), the intersections of CAs (left column) on the same horizontal slice, with a field of view identical to that in \figref{fig:swirl_show}. The snapshots shown in \figref{fig:Q_cluster_tw} cover the development of the twisted flux rope driven by the photospheric vortex. Both the $Q$ factor and the twist number show very complex structures, which actually map the complex current structures in the magnetic field of the quiet‑Sun corona. Therefore, it is not practical to isolate a single coherent structure, although areas of high twist that are cospatial with the core positive flux concentration can be clearly identified.

The CAs identified by our clustering method clearly demonstrate how the twisted flux rope forms. The intersections of the CAs are overlaid on the map of the $Q$ factor. The boundaries of different magnetic structures coincide with the regions of high $Q$ value (gray contours) and high $Q$ gradient (black lines), which correspond to QSLs. \figref{fig:qsl_interface} provides a quantitative comparison between the QSLs and the interfaces of the different CAs. The left panel displays a map of $Q$, which is overlaid with colored lines that represent CA interfaces. The interfaces between the \textit{flux rope} and other CAs are colored according to the designated colors of CAs; the interfaces among the other seven CAs (i.e., excluding the \textit{flux rope}) are shown in black.\footnote{The   \textit{ambient field} and \textit{flux rope} have no boundary within this field of view} Only a few CA interfaces appear as smooth and continuous curves; whereas most of them are fragmented segments or tiny closed loops. Nevertheless, a visual comparison already reveals a correlation between the CA interfaces and lines of large $Q$. The middle panel shows histograms of $Q$ values at points on the CA interfaces (blue) and those over the whole region (red). In average, the $Q$ values at the CA interfaces are larger than the mean values over the whole region by 3–4 orders of magnitude. Furthermore, to account for the circumstance that the CA interfaces are close but are not immediately overlaid QSLs, we evaluate $Q_{\mathrm{max}}$ by
\begin{equation}
Q_{\max}(\mathbf{x}_i,\epsilon)=\max_{\left| \mathbf{x}-\mathbf{x}_i \right| \leq \epsilon}
Q(\mathbf{x}),
\end{equation}
where $x_{i}$ is a pixel at the CA interfaces and $x$ are the points within a neighborhood of a radius $\epsilon$.

In the right panel, we compare the histograms of $Q$ values with $\epsilon=0.024$ and 0.192 Mm, and the latter is the native grid spacing of the simulation. It is clear that when given a tolerance of one grid point, $Q$ values on CA interfaces are significantly increase. These results confirm that our clustering method correctly identifies different magnetic structures, and that their boundary are located on, or very close to, regions where $Q$ attains very large magnitudes, i.e., QSLs. In the early stage (top row of \figref{fig:Q_cluster_tw}), the intersections of different magnetic structures are still relatively isolated. The animation of \figref{fig:Q_cluster_tw} shows that the counterclockwise rotation of the flux concentration continuously deforms the CAs. More specifically, the other four CAs that share the same positive concentration as \textit{flux rope} are strongly stretched and wrapped around the trunk of the flux rope. At the moment of the rapid eruption (bottom row of \figref{fig:Q_cluster_tw}), the footpoints of \textit{loop E} and \textit{loop SE} become deeply embedded within \textit{flux rope}, leading to a large twist number in the core positive flux concentration.

Small‑scale flux concentrations formed at converging downflow lanes of granulation are ubiquitous in the quiet Sun. Therefore, why a flux rope can form at this particular location remains unclear. We analyze the coupling between the vortex and the magnetic field on the horizontal slice at $z = 1.584$~Mm. We identify individual flux concentrations enclosed by green contours of $|B_z| = 50$~G, as illustrated in the left panel of \figref{fig:swirl_dist}. The right panel of \figref{fig:swirl_dist} shows the joint distribution of the vertical magnetic flux of the identified concentrations ($\phi$), given by
\begin{equation}
\phi = \int_{S} B_z \, dS,
\end{equation}
and the vertical vorticity flux over the same area ($\Gamma$), the latter is estimated by
\begin{equation}
\Gamma = \int_{S} \omega_z \, dS = \oint_{\partial S} \vec{v} \cdot d\vec{l}.
\end{equation}
We compute the full velocity circulation integral along the boundaries of all the identified concentrations.

The calculation is performed independently for each snapshot, and the data points are overlaid in the same plot with colors corresponding to time. The black line indicates the core positive flux concentration (i.e., the positive footpoint of \textit{flux rope}). In the early stage ($t < 30$ min), this core positive concentration has a noticeably larger magnetic flux than most quiet‑Sun flux concentrations do, whereas its vorticity flux is not significant and remains below $1.0 \times 10^{15}$ cm$^{2}$ s$^{-1}$. Both parameters steadily increase, and at $t = 51$ min, the flux rope footpoint becomes a unique point with the largest vorticity flux ($\Gamma \approx 2.8 \times 10^{15}$ cm$^{2}$ s$^{-1}$) coupled with a large magnetic flux ($\phi \approx 3.5 \times 10^{19}$\,Mx). During the rapid eruption stage, the lifting and unwinding motion of \textit{flux rope} contributes to clockwise rotation, which is opposite to the driving counterclockwise vortex and causes an apparent decrease in the vorticity flux.

\subsection{Deeper Root of the Photospheric Vortex}
\begin{figure*}
  \centering
  \includegraphics[width=18cm]{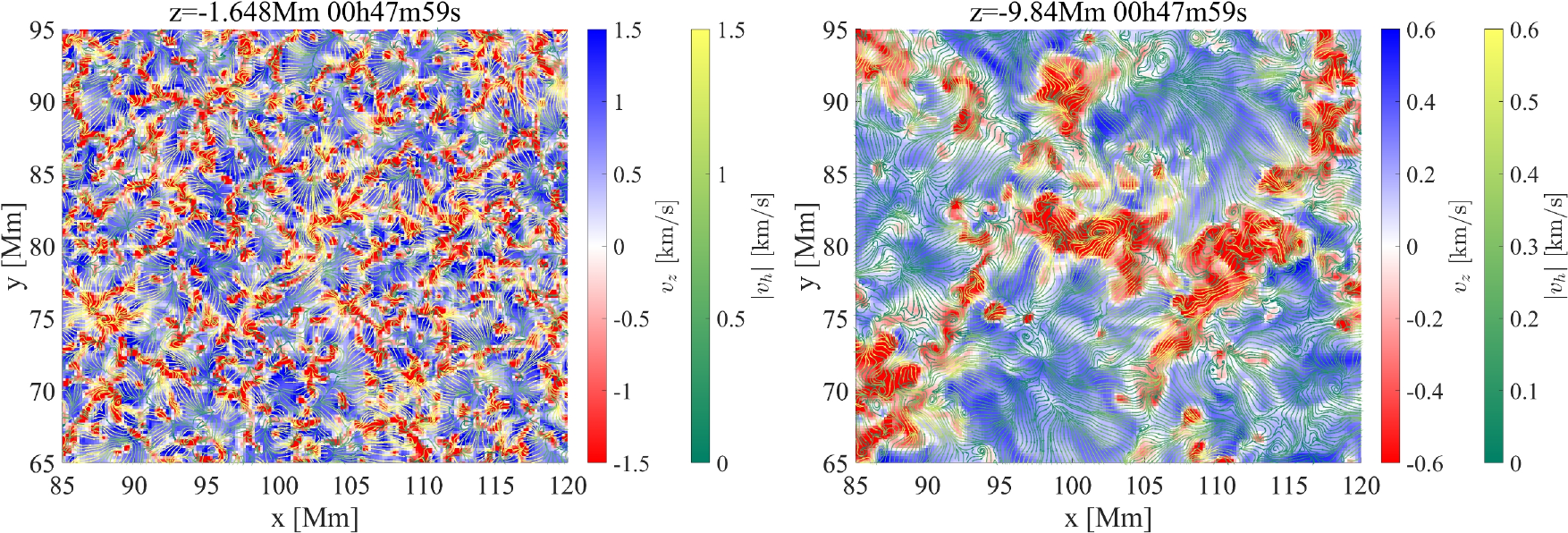}
\caption{Flow field in horizontal slices at $z = -1.648$~Mm and $z = -9.840$~Mm. The field of view is the same as that in panel (c) of \figref{fig:swirl_show}. An animation that covers the dynamics evolution of 72 minutes starting from $t=t_0$ is available.
(An animation of this figure is available.)}
  \label{fig:convection_zone}
\end{figure*}

Furthermore, we examine the origin of the vortex in the convective zone. \figref{fig:convection_zone} shows the flow field at two different depths over the same field of view as \figref{fig:swirl_show}. The horizontal flows are displayed as streamlines on the map of the vertical velocity.

In the shallower layer at $z=-1.648$\,Mm, the vertical flow field consists of upflow cells of a few Mm and thin downflow lanes at the boundaries of the upflows. The horizontal flow field presents many rotating and shear motions. The most evident vortex can be found near the center of the field of view, which is cospatial with the footpoint of \textit{flux rope}. The animation shows a continuous counterclock-wise rotation that drives small patches of downflows winding around the center of the swirl, which is consistent with the evolution of the photospheric magnetic flux concentrations and coronal magnetic field lines presented earlier in the paper.

At a greater depth of $z=-9.84$\,Mm, the scales of the convective motions become significantly larger. The concentrations of the downflow lanes are found as the boundary of broad upflow cells on scales of about 20\,Mm. At this depth, the lifetime of supergranular-scale motions is expected to be much longer than that of activities in the atmosphere; however, short-timescale dynamics may actively modulate the dynamics in shallower layers. The animation demonstrates that one hour before \textit{flux rope} eruption, the upflow cells around the center downflow lanes have been pushing the patches of downflows toward a converging point, and the streamlines of the horizontal flow field indicate clear swirl feeding into the center. Moreover, we examine the flow field down to a depth of about 20\,Mm below the surface (i.e., the bottom boundary of the simulation domain) and find a similar evolution.

This process gives rise to a vortex tube that extends from long-lifetime and larger-scale convective downflow to the solar atmosphere and drives the continuous swirling motion that twists coronal magnetic field lines. The deeper root of the downflow and the vortex appears to be crucial for the two ingredients for the formation of the flux rope —- a strong magnetic flux concentration and a persistent swirling motion.

\subsection{Observable Features of Flux Rope Eruption}
\label{subsect:Observable Features of the Flux Rope Eruption}
\begin{figure*}
 \centering
 \includegraphics[width=18cm]{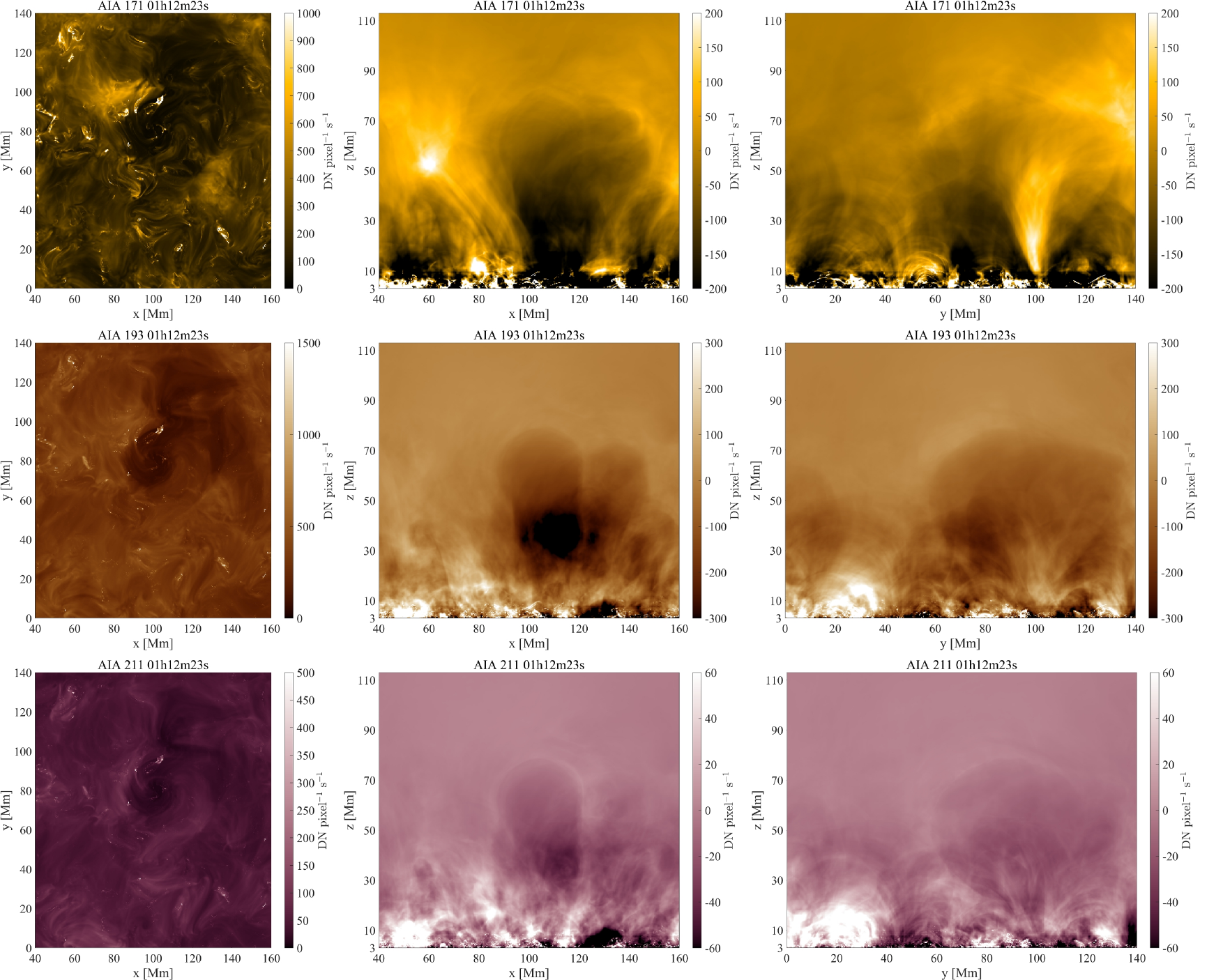}
 \caption{Synthetic EUV images of a subregion at AIA 171, 193, and 211 \,\AA~ channels. The first column shows the original images from the top view. The second and third columns show the images after subtracting the background. The field of view is the same as the black rectangle in \figref{fig:aia_magneto_cluster_twist}(a). An animation that covers the dynamics evolution of 72 minutes starting from $t=t_0$ is available. (An animation of this figure is available.)} 
 \label{fig:AIA_subreg}
\end{figure*}

To illustrate the observable signatures of the flux rope eruption, we synthesize the AIA 171, 193, and 211 \,\AA~ emission in the volume marked by the black rectangle in \figref{fig:aia_magneto_cluster_twist}(a). We cut off the emission below $z = 3$~Mm to highlight the coronal disturbance. \figref{fig:AIA_subreg} shows the dimming feature in the final eruptive phase. The left column displays the images integrated along the line of sight from the top, while the middle and right columns display the base difference images seen from two side views. The base difference is obtained by subtracting the horizontally averaged intensity \footnote{The horizontal average is performed over the entire horizontal domain, rather than the area marked by the black rectangle in \figref{fig:aia_magneto_cluster_twist}(a), to minimize the impact of individual bright or dark features on the base difference.} at $t_{0}$.

\textit{Flux rope} consistently exhibits a dimming region in all three channels, which covers a temperature range from slightly below 1~MK to above 2~MK. This implies that the dimming is not caused by cold plasma but by the expansion of the magnetic flux rope. The animation of \figref{fig:AIA_subreg} shows the observable trace of \textit{flux rope}. At $t = 29$~m~47~s, a balloon-shaped dimming appears and propagates in the $+x$ and $+z$ directions. At $t = 59$~m~9~s, another dimming component appears and rises quickly in the vertical direction. This phenomenon confirms that the flux rope eruption proceeds in two distinct phases: an initial slow expansion phase when \textit{flux rope} gradually forms followed by a rapid acceleration phase after the reconnection, as we illustrated in \figref{fig:twist_FR}.

As a side note, the bright region located around $(x,y) = (80,100)$ at 171\,\AA~ is part of a static coronal loop, which is not related to the eruptive region.

\section{Summary and Discussion}
\label{Summary and Discussion}

In this study, we investigate the self-consistent formation, evolution, and eruption of a magnetic flux rope in the large-domain quiet-Sun radiative-MHD simulation of \citet{Chen-2022ApJ...937...91C}, which extends from the upper convection zone to the corona. To identify coherent magnetic structures in the highly complex quiet Sun corona, we develop a clustering method based on magnetic field-line connectivity, which allows us to distinguish the evolving flux rope from neighboring magnetic field groups and follows their interactions through the event. 

We find that a persistent counterclockwise vortex, which is rooted in a converging downflow extending into the convection zone for more than 10 Mm, continuously drives rotation of a relatively large quiet Sun flux concentration and forces surrounding magnetic structures wrapping around the central flux bundle. As a result, a initially weakly-twisted magnetic loops evolves into a highly twisted flux rope and eventually erupts in a scenario similar to that proposed in active-region-scale flare and CME models \citep{Aulanier-2012A&A...543A.110A}. 

Previous studies have demonstrated that vortex motions (swirls) are ubiquitous in the quiet Sun and can modify the magnetic structures in the lower solar atmosphere\citep{Liu-2019ApJ...872...22L,Amari-2025ApJ...984L..37A,Xie-2025A&A...700A...6X,Xie-2025ApJ...979...27X}. The  helicity injection into the atmosphere through magnetized vortex tubes has also been investigated by \citet{Kitiashvili-2012ApJ...751L..21K,Kitiashvili-2013ApJ...770...37K}. Compared with previous simulations of quiet-Sun vortices \citep[see also, ][]{Kannan-2024MNRAS.533.3611K}, our model covers a considerably broad horizontal domain of about 200 Mm and a vertical extent from 18~Mm below the photosphere to 113~Mm in the corona, allowing us to follow the complete evolution from a deeper convective layer to coronal eruption. In particular, the long-lived converging downflow and vortex can develop on supergranular scales in this simulation. Such processes have also been observed in the real Sun \citet{Requerey-2018A&A...610A..84R,Chian-2019MNRAS.488.3076C}. Our simulation indicates that the coincidence of a persistent vortex and a large quiet-Sun flux concentration can create an intermediate flux rope on a scale different from commonly observed mini flux ropes in the quiet Sun or large ones in active regions.

Propagating dimmings similar to the observable signature of the \textit{flux rope} eruption have been reported in the quiet Sun. \citet{Jiang-2007SoPh..240...77J} attributed H$\alpha$ and EUV dimming associated with a coronal sigmoid eruption to a flux rope system. \citet{Wang-2023ApJ...952L..29W} analyzed a mini coronal dimming and linked it to a mini-filament eruption, with mass loss estimated from DEM analysis. \citet{Mou-2018A&A...619A..55M} reported that dimming accompanying mini-CME from coronal bright points arises from both cool ejected plasma and depleted coronal density. \citet{Yang-2015ApJ...803...86Y} reported dimming from a circular filament eruption. Our simulated dimming is consistent with these observations in terms of originating from an erupting magnetic structure, but its physical cause differs: it is driven primarily by the expansion of an initially low-density flux rope, rather than by mass loss of a plasma-hosting flux rope or filament evacuation. The dimming spans about 40 Mm horizontally and up to 100 Mm vertically, comparable in scale to coronal cavities that were found in polarimetric observations and suggested to be signatures of coronal flux ropes \citep[e.g.,][]{BakSteslicka-2013ApJ...770L..28B,Ruminska-2022ApJ...926..146R}.

Several minor processes accompany the main eruption, though they are not the primary drivers of flux rope formation. Weak flux emergence beneath \textit{filament S} exhibits a serpentine morphology reminiscent of the undulatory flux tubes described by \citet{Pariat-2004ApJ...614.1099P,Pariat-2006AdSpR..38..902P,Pariat-2009ApJ...701.1911P}. In addition, a negative-polarity patch intrudes into the core positive flux concentration, similar to the intrusive patterns identified in active regions \citep{Yang-2017ApJ...849L..21Y,Bamba-2020ApJ...894...29B,Inoue-2021ApJ...914...71I}. However, in our quiet-Sun case, this intrusion stretches and rotates \textit{filament S} and pushes cool plasma into the corona. The persistent boundary between \textit{flux rope} and \textit{filament S} throughout the evolution (Fig. \ref{fig:Q_cluster_tw}) is topologically analogous to so-called sunspot scars identified in active regions \citep{Xing-2024A&A...682A...3X}, which mark the edge of a flux rope footpoint—the separatrix between the flux rope and surrounding loops. This suggests that such boundaries are not exclusive to large sunspots in active regions, but can exist wherever a flux rope and a distinct adjacent magnetic system coexist, even on the much smaller scales of the quiet Sun.

To conclude, our simulation demonstrates that a persistent vortex rooted in greater depth beneath the solar surface can progressively twist and reorganize quiet-Sun magnetic fields, creating a coherent flux rope that eventually erupts through a process analogous to large eruptions in active regions. Although the event develops current sheets and localized heating, the dominant observable signature is an EUV dimming caused primarily by density rarefaction. Such stealth-like eruptions may therefore be more common in real Sun than previously recognized and could become increasingly detectable with future observations of higher sensitivity and spatial resolution.

\begin{acknowledgments}
We thank the referee for constructive suggestions that improved the presentation of the results. The authors thank Drs. Chen Xing, Zekun Lu, Can Wang, Jinhan Guo, Yang Guo, Xin Cheng and Yulei Wang for helpful discussions. This work is supported by National Science Foundation of China No. 12422308 and No. 12373054.
\end{acknowledgments}

\bibliography{references}{}

@ARTICLE{Priest-2002A&ARv..10..313P,
       author = {{Priest}, E.~R. and {Forbes}, T.~G.},
        title = "{The magnetic nature of solar flares}",
      journal = {\aapr},
         year = 2002,
        month = jan,
       volume = {10},
       number = {4},
        pages = {313-377},
          doi = {10.1007/s001590100013},
       adsurl = {https://ui.adsabs.harvard.edu/abs/2002A&ARv..10..313P}
}

@ARTICLE{Liu-2020RAA....20..165L,
       author = {{Liu}, Rui},
        title = "{Magnetic flux ropes in the solar corona: structure and evolution toward eruption}",
      journal = {Research in Astronomy and Astrophysics},
         year = 2020,
        month = oct,
       volume = {20},
       number = {10},
          eid = {165},
        pages = {165},
          doi = {10.1088/1674-4527/20/10/165},
archivePrefix = {arXiv},
       eprint = {2007.11363},
 primaryClass = {astro-ph.SR},
       adsurl = {https://ui.adsabs.harvard.edu/abs/2020RAA....20..165L}
}

@ARTICLE{Cheng-2020ApJ...894...85C,
       author = {{Cheng}, X. and {Zhang}, J. and {Kliem}, B. and {T{\"o}r{\"o}k}, T. and {Xing}, C. and {Zhou}, Z.~J. and {Inhester}, B. and {Ding}, M.~D.},
        title = "{Initiation and Early Kinematic Evolution of Solar Eruptions}",
      journal = {\apj},
         year = 2020,
        month = may,
       volume = {894},
       number = {2},
          eid = {85},
        pages = {85},
          doi = {10.3847/1538-4357/ab886a},
archivePrefix = {arXiv},
       eprint = {2004.03790},
 primaryClass = {astro-ph.SR},
       adsurl = {https://ui.adsabs.harvard.edu/abs/2020ApJ...894...85C}
}

@ARTICLE{Aulanier-2012A&A...543A.110A,
       author = {{Aulanier}, G. and {Janvier}, M. and {Schmieder}, B.},
        title = "{The standard flare model in three dimensions. I. Strong-to-weak shear transition in post-flare loops}",
      journal = {\aap},
         year = 2012,
        month = jul,
       volume = {543},
          eid = {A110},
        pages = {A110},
          doi = {10.1051/0004-6361/201219311},
       adsurl = {https://ui.adsabs.harvard.edu/abs/2012A&A...543A.110A}
}

@ARTICLE{Janvier-2013A&A...555A..77J,
       author = {{Janvier}, M. and {Aulanier}, G. and {Pariat}, E. and {D{\'e}moulin}, P.},
        title = "{The standard flare model in three dimensions. III. Slip-running reconnection properties}",
      journal = {\aap},
         year = 2013,
        month = jul,
       volume = {555},
          eid = {A77},
        pages = {A77},
          doi = {10.1051/0004-6361/201321164},
archivePrefix = {arXiv},
       eprint = {1305.4053},
 primaryClass = {astro-ph.SR},
       adsurl = {https://ui.adsabs.harvard.edu/abs/2013A&A...555A..77J}
}

@ARTICLE{Leka-1996ApJ...462..547L,
       author = {{Leka}, K.~D. and {Canfield}, R.~C. and {McClymont}, A.~N. and {van Driel-Gesztelyi}, L.},
        title = "{Evidence for Current-carrying Emerging Flux}",
      journal = {\apj},
         year = 1996,
        month = may,
       volume = {462},
        pages = {547},
          doi = {10.1086/177171},
       adsurl = {https://ui.adsabs.harvard.edu/abs/1996ApJ...462..547L}
}

@ARTICLE{Fan-2001ApJ...554L.111F,
       author = {{Fan}, Y.},
        title = "{The Emergence of a Twisted {\ensuremath{\Omega}}-Tube into the Solar Atmosphere}",
      journal = {\apjl},
         year = 2001,
        month = jun,
       volume = {554},
       number = {1},
        pages = {L111-L114},
          doi = {10.1086/320935},
       adsurl = {https://ui.adsabs.harvard.edu/abs/2001ApJ...554L.111F}
}

@INPROCEEDINGS{Torok-2001sps..proc..364T,
       author = {{T{\"o}r{\"o}k}, T. and {Kliem}, B.},
        title = "{The evolution of coronal magnetic flux tubes under the influence of footpoint twisting motions}",
    booktitle = {Space Plasma Simulation},
         year = 2001,
       editor = {{B{\"u}chner}, J. and {Dum}, C.~T. and {Scholer}, M.},
        month = jan,
        pages = {364},
       adsurl = {https://ui.adsabs.harvard.edu/abs/2001sps..proc..364T}
}

@ARTICLE{Ballegooijen-1989ApJ...343..971V,
       author = {{van Ballegooijen}, A.~A. and {Martens}, P.~C.~H.},
        title = "{Formation and Eruption of Solar Prominences}",
      journal = {\apj},
         year = 1989,
        month = aug,
       volume = {343},
        pages = {971},
          doi = {10.1086/167766},
       adsurl = {https://ui.adsabs.harvard.edu/abs/1989ApJ...343..971V}
}

@ARTICLE{Klimchuk-1992ApJ...385..344K,
       author = {{Klimchuk}, J.~A. and {Sturrock}, P.~A.},
        title = "{Three-dimensional Force-free Magnetic Fields and Flare Energy Buildup}",
      journal = {\apj},
         year = 1992,
        month = jan,
       volume = {385},
        pages = {344},
          doi = {10.1086/170943},
       adsurl = {https://ui.adsabs.harvard.edu/abs/1992ApJ...385..344K}
}

@ARTICLE{Balmaceda-2010A&A...513L...6B,
       author = {{Balmaceda}, L. and {Vargas Dom{\'\i}nguez}, S. and {Palacios}, J. and {Cabello}, I. and {Domingo}, V.},
        title = "{Evidence of small-scale magnetic concentrations dragged by vortex motion of solar photospheric plasma}",
      journal = {\aap},
         year = 2010,
        month = apr,
       volume = {513},
          eid = {L6},
        pages = {L6},
          doi = {10.1051/0004-6361/200913584},
archivePrefix = {arXiv},
       eprint = {1004.1185},
 primaryClass = {astro-ph.SR},
       adsurl = {https://ui.adsabs.harvard.edu/abs/2010A&A...513L...6B}
}

@ARTICLE{Requerey-2017ApJS..229...14R,
       author = {{Requerey}, Iker S. and {Del Toro Iniesta}, Jose Carlos and {Bellot Rubio}, Luis R. and {Mart{\'\i}nez Pillet}, Valent{\'\i}n and {Solanki}, Sami K. and {Schmidt}, Wolfgang},
        title = "{Convectively Driven Sinks and Magnetic Fields in the Quiet-Sun}",
      journal = {\apjs},
         year = 2017,
        month = mar,
       volume = {229},
       number = {1},
          eid = {14},
        pages = {14},
          doi = {10.3847/1538-4365/229/1/14},
archivePrefix = {arXiv},
       eprint = {1610.07622},
 primaryClass = {astro-ph.SR},
       adsurl = {https://ui.adsabs.harvard.edu/abs/2017ApJS..229...14R}
}

@ARTICLE{Requerey-2018A&A...610A..84R,
       author = {{Requerey}, Iker S. and {Cobo}, Basilio Ruiz and {Go{\v{s}}i{\'c}}, Milan and {Bellot Rubio}, Luis R.},
        title = "{Persistent magnetic vortex flow at a supergranular vertex}",
      journal = {\aap},
         year = 2018,
        month = mar,
       volume = {610},
          eid = {A84},
        pages = {A84},
          doi = {10.1051/0004-6361/201731842},
archivePrefix = {arXiv},
       eprint = {1712.01510},
 primaryClass = {astro-ph.SR},
       adsurl = {https://ui.adsabs.harvard.edu/abs/2018A&A...610A..84R}
}

@ARTICLE{Chian-2019MNRAS.488.3076C,
       author = {{Chian}, Abraham C.-L. and {Silva}, Suzana S.~A. and {Rempel}, Erico L. and {Go{\v{s}}i{\'c}} and {}, Milan and {Bellot Rubio}, Luis R. and {Kusano}, Kanya and {Miranda}, Rodrigo A. and {Requerey}, Iker S.},
        title = "{Supergranular turbulence in the quiet Sun: Lagrangian coherent structures}",
      journal = {\mnras},
         year = 2019,
        month = sep,
       volume = {488},
       number = {3},
        pages = {3076-3088},
          doi = {10.1093/mnras/stz1909},
archivePrefix = {arXiv},
       eprint = {1904.08472},
 primaryClass = {astro-ph.SR},
       adsurl = {https://ui.adsabs.harvard.edu/abs/2019MNRAS.488.3076C}
}

@ARTICLE{Murabito-2020A&A...639A..59M,
       author = {{Murabito}, Mariarita and {Shetye}, Juie and {Stangalini}, Marco and {Verwichte}, Erwin and {Arber}, Tony and {Ermolli}, Ilaria and {Giorgi}, Fabrizio and {Goffrey}, Tom},
        title = "{Unveiling the magnetic nature of chromospheric vortices}",
      journal = {\aap},
         year = 2020,
        month = jul,
       volume = {639},
          eid = {A59},
        pages = {A59},
          doi = {10.1051/0004-6361/202038360},
archivePrefix = {arXiv},
       eprint = {2006.13776},
 primaryClass = {astro-ph.SR},
       adsurl = {https://ui.adsabs.harvard.edu/abs/2020A&A...639A..59M}
}

@ARTICLE{Tziotziou-2023SSRv..219....1T,
       author = {{Tziotziou}, K. and {Scullion}, E. and {Shelyag}, S. and {Steiner}, O. and {Khomenko}, E. and {Tsiropoula}, G. and {Canivete Cuissa}, J.~R. and {Wedemeyer}, S. and {Kontogiannis}, I. and {Yadav}, N. and et al.},
        title = "{Vortex Motions in the Solar Atmosphere}",
      journal = {\ssr},
         year = 2023,
        month = feb,
       volume = {219},
       number = {1},
          eid = {1},
        pages = {1},
          doi = {10.1007/s11214-022-00946-8},
       adsurl = {https://ui.adsabs.harvard.edu/abs/2023SSRv..219....1T}
}

@ARTICLE{Yang-2015ApJ...803...86Y,
       author = {{Yang}, Bo and {Jiang}, Yunchun and {Yang}, Jiayan and {Hong}, Junchao and {Xu}, Zhe},
        title = "{The Formation and Eruption of a Small Circular Filament Driven by Rotating Magnetic Structures in the Quiet Sun}",
      journal = {\apj},
         year = 2015,
        month = apr,
       volume = {803},
       number = {2},
          eid = {86},
        pages = {86},
          doi = {10.1088/0004-637X/803/2/86},
       adsurl = {https://ui.adsabs.harvard.edu/abs/2015ApJ...803...86Y}
}

@ARTICLE{Dai-2021ApJ...906...66D,
       author = {{Dai}, Jun and {Ji}, Haisheng and {Li}, Leping and {Zhang}, Jun and {Chen}, Huadong},
        title = "{The Formation and Eruption of a Sigmoidal Filament Driven by Rotating Network Magnetic Fields}",
      journal = {\apj},
         year = 2021,
        month = jan,
       volume = {906},
       number = {2},
          eid = {66},
        pages = {66},
          doi = {10.3847/1538-4357/abcaf4},
archivePrefix = {arXiv},
       eprint = {2012.06775},
 primaryClass = {astro-ph.SR},
       adsurl = {https://ui.adsabs.harvard.edu/abs/2021ApJ...906...66D}
}

@ARTICLE{Brown-2003SoPh..216...79B,
       author = {{Brown}, D.~S. and {Nightingale}, R.~W. and {Alexander}, D. and {Schrijver}, C.~J. and {Metcalf}, T.~R. and {Shine}, R.~A. and {Title}, A.~M. and {Wolfson}, C.~J.},
        title = "{Observations of Rotating Sunspots from TRACE}",
      journal = {\solphys},
         year = 2003,
        month = sep,
       volume = {216},
       number = {1},
        pages = {79-108},
          doi = {10.1023/A:1026138413791},
       adsurl = {https://ui.adsabs.harvard.edu/abs/2003SoPh..216...79B}
}

@ARTICLE{Vemareddy-2012ApJ...761...60V,
       author = {{Vemareddy}, P. and {Ambastha}, A. and {Maurya}, R.~A.},
        title = "{On the Role of Rotating Sunspots in the Activity of Solar Active Region NOAA 11158}",
      journal = {\apj},
         year = 2012,
        month = dec,
       volume = {761},
       number = {1},
          eid = {60},
        pages = {60},
          doi = {10.1088/0004-637X/761/1/60},
archivePrefix = {arXiv},
       eprint = {1210.3912},
 primaryClass = {astro-ph.SR},
       adsurl = {https://ui.adsabs.harvard.edu/abs/2012ApJ...761...60V}
}

@ARTICLE{Ruan-2014ApJ...784..165R,
       author = {{Ruan}, Guiping and {Chen}, Yao and {Wang}, Shuo and {Zhang}, Hongqi and {Li}, Gang and {Jing}, Ju and {Su}, Jiangtao and {Li}, Xing and {Xu}, Haiqing and {Du}, Guohui and et al.},
        title = "{A Solar Eruption Driven by Rapid Sunspot Rotation}",
      journal = {\apj},
         year = 2014,
        month = apr,
       volume = {784},
       number = {2},
          eid = {165},
        pages = {165},
          doi = {10.1088/0004-637X/784/2/165},
archivePrefix = {arXiv},
       eprint = {1402.6043},
 primaryClass = {astro-ph.SR},
       adsurl = {https://ui.adsabs.harvard.edu/abs/2014ApJ...784..165R}
}

@ARTICLE{Zhang-2022ApJ...933..200Z,
       author = {{Zhang}, Yan and {Yan}, Xiaoli and {Wang}, Jincheng and {Li}, Qiaoling and {Yang}, Liheng and {Xue}, Zhike},
        title = "{A Double-decker Filament Formation Driven by Sunspot Rotation and Magnetic Reconnection}",
      journal = {\apj},
         year = 2022,
        month = jul,
       volume = {933},
       number = {2},
          eid = {200},
        pages = {200},
          doi = {10.3847/1538-4357/ac7391},
       adsurl = {https://ui.adsabs.harvard.edu/abs/2022ApJ...933..200Z}
}

@ARTICLE{Wedemeyer-2012Natur.486..505W,
       author = {{Wedemeyer-B{\"o}hm}, Sven and {Scullion}, Eamon and {Steiner}, Oskar and {Rouppe van der Voort}, Luc and {de La Cruz Rodriguez}, Jaime and {Fedun}, Viktor and {Erd{\'e}lyi}, Robert},
        title = "{Magnetic tornadoes as energy channels into the solar corona}",
      journal = {\nat},
         year = 2012,
        month = jun,
       volume = {486},
       number = {7404},
        pages = {505-508},
          doi = {10.1038/nature11202},
       adsurl = {https://ui.adsabs.harvard.edu/abs/2012Natur.486..505W}
}

@ARTICLE{Huang-2019ApJ...887..130H,
       author = {{Huang}, Z.~W. and {Cheng}, X. and {Su}, Y.~N. and {Liu}, T. and {Ding}, M.~D.},
        title = "{Formation and Eruption of a Mini-sigmoid Originating in Coronal Hole}",
      journal = {\apj},
         year = 2019,
        month = dec,
       volume = {887},
       number = {2},
          eid = {130},
        pages = {130},
          doi = {10.3847/1538-4357/ab4f83},
archivePrefix = {arXiv},
       eprint = {1912.10404},
 primaryClass = {astro-ph.SR},
       adsurl = {https://ui.adsabs.harvard.edu/abs/2019ApJ...887..130H}
}

@ARTICLE{Tziotziou-2018A&A...618A..51T,
       author = {{Tziotziou}, K. and {Tsiropoula}, G. and {Kontogiannis}, I. and {Scullion}, E. and {Doyle}, J.~G.},
        title = "{A persistent quiet-Sun small-scale tornado. I. Characteristics and dynamics}",
      journal = {\aap},
         year = 2018,
        month = oct,
       volume = {618},
          eid = {A51},
        pages = {A51},
          doi = {10.1051/0004-6361/201833101},
       adsurl = {https://ui.adsabs.harvard.edu/abs/2018A&A...618A..51T}
}

@ARTICLE{Silva-2020ApJ...898..137S,
       author = {{Silva}, Suzana S.~A. and {Fedun}, Viktor and {Verth}, Gary and {Rempel}, Erico L. and {Shelyag}, Sergiy},
        title = "{Solar Vortex Tubes: Vortex Dynamics in the Solar Atmosphere}",
      journal = {\apj},
         year = 2020,
        month = aug,
       volume = {898},
       number = {2},
          eid = {137},
        pages = {137},
          doi = {10.3847/1538-4357/ab99a9},
archivePrefix = {arXiv},
       eprint = {2007.04371},
 primaryClass = {astro-ph.SR},
       adsurl = {https://ui.adsabs.harvard.edu/abs/2020ApJ...898..137S}
}

@ARTICLE{Silva-2021ApJ...915...24S,
       author = {{Silva}, Suzana S.~A. and {Verth}, Gary and {Rempel}, Erico L. and {Shelyag}, Sergiy and {Schiavo}, Luiz A.~C.~A. and {Fedun}, Viktor},
        title = "{Solar Vortex Tubes. II. On the Origin of Magnetic Vortices}",
      journal = {\apj},
         year = 2021,
        month = jul,
       volume = {915},
       number = {1},
          eid = {24},
        pages = {24},
          doi = {10.3847/1538-4357/abfec2},
       adsurl = {https://ui.adsabs.harvard.edu/abs/2021ApJ...915...24S}
}

@ARTICLE{Silva-2024ApJ...975..118S,
       author = {{Silva}, Suzana S.~A. and {Verth}, Gary and {Ballai}, Istvan and {Rempel}, Erico L. and {Shelyag}, Sergiy and {Schiavo}, Luiz A.~C.~A. and {Gomes}, Tiago F.~P. and {Fedun}, Viktor},
        title = "{Solar Vortex Tubes. III. Vorticity and Energy Transport}",
      journal = {\apj},
         year = 2024,
        month = nov,
       volume = {975},
       number = {1},
          eid = {118},
        pages = {118},
          doi = {10.3847/1538-4357/ad781a},
       adsurl = {https://ui.adsabs.harvard.edu/abs/2024ApJ...975..118S}
}

@ARTICLE{Battaglia-2021A&A...649A.121B,
       author = {{Battaglia}, Andrea Francesco and {Canivete Cuissa}, Jos{\'e} Roberto and {Calvo}, Flavio and {Bossart}, Aleksi Antoine and {Steiner}, Oskar},
        title = "{The Alfv{\'e}nic nature of chromospheric swirls}",
      journal = {\aap},
         year = 2021,
        month = may,
       volume = {649},
          eid = {A121},
        pages = {A121},
          doi = {10.1051/0004-6361/202040110},
archivePrefix = {arXiv},
       eprint = {2103.07366},
 primaryClass = {astro-ph.SR},
       adsurl = {https://ui.adsabs.harvard.edu/abs/2021A&A...649A.121B}
}

@ARTICLE{Chen-2022ApJ...937...91C,
       author = {{Chen}, Feng and {Rempel}, Matthias and {Fan}, Yuhong},
        title = "{A Comprehensive Radiative Magnetohydrodynamics Simulation of Active Region Scale Flux Emergence from the Convection Zone to the Corona}",
      journal = {\apj},
         year = 2022,
        month = oct,
       volume = {937},
       number = {2},
          eid = {91},
        pages = {91},
          doi = {10.3847/1538-4357/ac8f95},
archivePrefix = {arXiv},
       eprint = {2106.14055},
 primaryClass = {astro-ph.SR},
       adsurl = {https://ui.adsabs.harvard.edu/abs/2022ApJ...937...91C}
}

@ARTICLE{Titov-2002JGRA..107.1164T,
       author = {{Titov}, Vyacheslav S. and {Hornig}, Gunnar and {D{\'e}moulin}, Pascal},
        title = "{Theory of magnetic connectivity in the solar corona}",
      journal = {Journal of Geophysical Research (Space Physics)},
         year = 2002,
        month = aug,
       volume = {107},
       number = {A8},
          eid = {1164},
        pages = {1164},
          doi = {10.1029/2001JA000278},
       adsurl = {https://ui.adsabs.harvard.edu/abs/2002JGRA..107.1164T}
}

@ARTICLE{Titov-2007ApJ...660..863T,
       author = {{Titov}, V.~S.},
        title = "{Generalized Squashing Factors for Covariant Description of Magnetic Connectivity in the Solar Corona}",
      journal = {\apj},
         year = 2007,
        month = may,
       volume = {660},
       number = {1},
        pages = {863-873},
          doi = {10.1086/512671},
archivePrefix = {arXiv},
       eprint = {astro-ph/0703671},
 primaryClass = {astro-ph},
       adsurl = {https://ui.adsabs.harvard.edu/abs/2007ApJ...660..863T}
}

@ARTICLE{Berger-2006JPhA...39.8321B,
       author = {{Berger}, Mitchell A. and {Prior}, Chris},
        title = "{The writhe of open and closed curves}",
      journal = {Journal of Physics A Mathematical General},
         year = 2006,
        month = jun,
       volume = {39},
       number = {26},
        pages = {8321-8348},
          doi = {10.1088/0305-4470/39/26/005},
       adsurl = {https://ui.adsabs.harvard.edu/abs/2006JPhA...39.8321B}
}

@ARTICLE{Liu-2016ApJ...818..148L,
       author = {{Liu}, Rui and {Kliem}, Bernhard and {Titov}, Viacheslav S. and {Chen}, Jun and {Wang}, Yuming and {Wang}, Haimin and {Liu}, Chang and {Xu}, Yan and {Wiegelmann}, Thomas},
        title = "{Structure, Stability, and Evolution of Magnetic Flux Ropes from the Perspective of Magnetic Twist}",
      journal = {\apj},
         year = 2016,
        month = feb,
       volume = {818},
       number = {2},
          eid = {148},
        pages = {148},
          doi = {10.3847/0004-637X/818/2/148},
archivePrefix = {arXiv},
       eprint = {1512.02338},
 primaryClass = {astro-ph.SR},
       adsurl = {https://ui.adsabs.harvard.edu/abs/2016ApJ...818..148L}
}

@misc{Yang-2024zndo..13831036Y,
       author = {{Yang}, Kai E and {Xia}, Chun and {njuguoyang}},
        title = "{Kai-E-Yang/QSL: QSL v1}",
         year = 2024,
        month = sep,
          eid = {10.5281/zenodo.13831036},
          doi = {10.5281/zenodo.13831036},
      version = {v1},
    publisher = {Zenodo},
       adsurl = {https://ui.adsabs.harvard.edu/abs/2024zndo..13831036Y}
}

@ARTICLE{ChenJun-2022ApJ...937...26Z,
       author = {{Zhang}, PeiJin and {Chen}, Jun and {Liu}, Rui and {Wang}, ChuanBing},
        title = "{FastQSL: A Fast Computation Method for Quasi-separatrix Layers}",
      journal = {\apj},
         year = 2022,
        month = sep,
       volume = {937},
       number = {1},
          eid = {26},
        pages = {26},
          doi = {10.3847/1538-4357/ac8d61},
archivePrefix = {arXiv},
       eprint = {2208.12569},
 primaryClass = {astro-ph.SR},
       adsurl = {https://ui.adsabs.harvard.edu/abs/2022ApJ...937...26Z}
}

@ARTICLE{Scott-2017ApJ...848..117S,
       author = {{Scott}, Roger B. and {Pontin}, David I. and {Hornig}, Gunnar},
        title = "{On the Magnetic Squashing Factor and the Lie Transport of Tangents}",
      journal = {\apj},
         year = 2017,
        month = oct,
       volume = {848},
       number = {2},
          eid = {117},
        pages = {117},
          doi = {10.3847/1538-4357/aa8a64},
       adsurl = {https://ui.adsabs.harvard.edu/abs/2017ApJ...848..117S}
}

@ARTICLE{Tassev-2017ApJ...840...89T,
       author = {{Tassev}, Svetlin and {Savcheva}, Antonia},
        title = "{QSL Squasher: A Fast Quasi-separatrix Layer Map Calculator}",
      journal = {\apj},
         year = 2017,
        month = may,
       volume = {840},
       number = {2},
          eid = {89},
        pages = {89},
          doi = {10.3847/1538-4357/aa6f06},
archivePrefix = {arXiv},
       eprint = {1609.00724},
 primaryClass = {astro-ph.SR},
       adsurl = {https://ui.adsabs.harvard.edu/abs/2017ApJ...840...89T}
}

@ARTICLE{ML_Chen-2025ApJS..280...70C,
       author = {{Chen}, R. and {Lu}, W. and {Hao}, Q. and {Meng}, Y. and {Chen}, P.~F. and {Shi}, C.},
        title = "{Statistical Analyses of Solar Active Regions in SDO/HMI Magnetograms Detected by the Unsupervised Machine Learning Method DSARD}",
      journal = {\apjs},
         year = 2025,
        month = oct,
       volume = {280},
       number = {2},
          eid = {70},
        pages = {70},
          doi = {10.3847/1538-4365/adfdd8},
archivePrefix = {arXiv},
       eprint = {2502.17990},
 primaryClass = {astro-ph.SR},
       adsurl = {https://ui.adsabs.harvard.edu/abs/2025ApJS..280...70C}
}

@ARTICLE{ML_Caballero-2013SoPh..283..691C,
       author = {{Caballero}, C. and {Aranda}, M.~C.},
        title = "{A Comparative Study of Clustering Methods for Active Region Detection in Solar EUV Images}",
      journal = {\solphys},
         year = 2013,
        month = apr,
       volume = {283},
       number = {2},
        pages = {691-717},
          doi = {10.1007/s11207-013-0239-2},
       adsurl = {https://ui.adsabs.harvard.edu/abs/2013SoPh..283..691C}
}

@inproceedings{campello2013density,
  title={Density-Based Clustering Based on Hierarchical Density Estimates},
  author={Campello, Ricardo J. G. B. and Moulavi, Davoud and Sander, J},
  booktitle={Advances in Knowledge Discovery and Data Mining (PAKDD 2013)},
  pages={160--172},
  year={2013},
  volume={7819},
  series={Lecture Notes in Computer Science},
  publisher={Springer},
  doi={10.1007/978-3-642-37456-2_14}
}

@article{mcinnes2017hdbscan,
  title={hdbscan: Hierarchical density based clustering},
  author={McInnes, Leland and Healy, John and Astels, Steve},
  journal={Journal of Open Source Software},
  volume={2},
  number={11},
  pages={205},
  year={2017},
  doi={10.21105/joss.00205}
}

@ARTICLE{Pariat-2004ApJ...614.1099P,
       author = {{Pariat}, E. and {Aulanier}, G. and {Schmieder}, B. and {Georgoulis}, M.~K. and {Rust}, D.~M. and {Bernasconi}, P.~N.},
        title = "{Resistive Emergence of Undulatory Flux Tubes}",
      journal = {\apj},
         year = 2004,
        month = oct,
       volume = {614},
       number = {2},
        pages = {1099-1112},
          doi = {10.1086/423891},
       adsurl = {https://ui.adsabs.harvard.edu/abs/2004ApJ...614.1099P}
}

@ARTICLE{Pariat-2009ApJ...701.1911P,
       author = {{Pariat}, E. and {Masson}, S. and {Aulanier}, G.},
        title = "{Current Buildup in Emerging Serpentine Flux Tubes}",
      journal = {\apj},
         year = 2009,
        month = aug,
       volume = {701},
       number = {2},
        pages = {1911-1921},
          doi = {10.1088/0004-637X/701/2/1911},
       adsurl = {https://ui.adsabs.harvard.edu/abs/2009ApJ...701.1911P}
}

@ARTICLE{Pariat-2006AdSpR..38..902P,
       author = {{Pariat}, E. and {Aulanier}, G. and {Schmieder}, B. and {Georgoulis}, M.~K. and {Rust}, D.~M. and {Bernasconi}, P.~N.},
        title = "{Emergence of undulatory magnetic flux tubes by small scale reconnections}",
      journal = {Advances in Space Research},
         year = 2006,
        month = jan,
       volume = {38},
       number = {5},
        pages = {902-905},
          doi = {10.1016/j.asr.2006.03.040},
       adsurl = {https://ui.adsabs.harvard.edu/abs/2006AdSpR..38..902P}
}

@ARTICLE{Zhang-2011ApJ...741L...7Z,
       author = {{Zhang}, Jun and {Liu}, Yang},
        title = "{Ubiquitous Rotating Network Magnetic Fields and Extreme-ultraviolet Cyclones in the Quiet Sun}",
      journal = {\apjl},
         year = 2011,
        month = nov,
       volume = {741},
       number = {1},
          eid = {L7},
        pages = {L7},
          doi = {10.1088/2041-8205/741/1/L7},
archivePrefix = {arXiv},
       eprint = {1109.5003},
 primaryClass = {astro-ph.SR},
       adsurl = {https://ui.adsabs.harvard.edu/abs/2011ApJ...741L...7Z}
}

@ARTICLE{Yang-2017ApJ...849L..21Y,
       author = {{Yang}, Shuhong and {Zhang}, Jun and {Zhu}, Xiaoshuai and {Song}, Qiao},
        title = "{Block-induced Complex Structures Building the Flare-productive Solar Active Region 12673}",
      journal = {\apjl},
         year = 2017,
        month = nov,
       volume = {849},
       number = {2},
          eid = {L21},
        pages = {L21},
          doi = {10.3847/2041-8213/aa9476},
archivePrefix = {arXiv},
       eprint = {1710.06545},
 primaryClass = {astro-ph.SR},
       adsurl = {https://ui.adsabs.harvard.edu/abs/2017ApJ...849L..21Y}
}

@ARTICLE{Bamba-2020ApJ...894...29B,
       author = {{Bamba}, Yumi and {Inoue}, Satoshi and {Imada}, Shinsuke},
        title = "{Intrusion of Magnetic Peninsula toward the Neighboring Opposite-polarity Region That Triggers the Largest Solar Flare in Solar Cycle 24}",
      journal = {\apj},
         year = 2020,
        month = may,
       volume = {894},
       number = {1},
          eid = {29},
        pages = {29},
          doi = {10.3847/1538-4357/ab85ca},
archivePrefix = {arXiv},
       eprint = {2005.00688},
 primaryClass = {astro-ph.SR},
       adsurl = {https://ui.adsabs.harvard.edu/abs/2020ApJ...894...29B}
}

@ARTICLE{Inoue-2021ApJ...914...71I,
       author = {{Inoue}, Satoshi and {Bamba}, Yumi},
        title = "{An MHD Modeling of the Successive X2.2 and X9.3 Solar Flares of 2017 September 6}",
      journal = {\apj},
         year = 2021,
        month = jun,
       volume = {914},
       number = {1},
          eid = {71},
        pages = {71},
          doi = {10.3847/1538-4357/abf835},
archivePrefix = {arXiv},
       eprint = {2104.06639},
 primaryClass = {astro-ph.SR},
       adsurl = {https://ui.adsabs.harvard.edu/abs/2021ApJ...914...71I}
}

@ARTICLE{Canny-1986ITPAM...8..679C,
       author = {{Canny}, John},
        title = "{A Computational Approach to Edge Detection}",
      journal = {IEEE Transactions on Pattern Analysis and Machine Intelligence},
         year = 1986,
        month = nov,
       volume = {8},
       number = {6},
        pages = {679-698},
          doi = {10.1109/TPAMI.1986.4767851},
       adsurl = {https://ui.adsabs.harvard.edu/abs/1986ITPAM...8..679C}
}

@ARTICLE{Wang-2023ApJ...952L..29W,
       author = {{Wang}, Rui and {Liu}, Ying D. and {Zhao}, Xiaowei and {Hu}, Huidong},
        title = "{Observations of Mini Coronal Dimmings Caused by Small-scale Eruptions in the Quiet Sun}",
      journal = {\apjl},
         year = 2023,
        month = aug,
       volume = {952},
       number = {2},
          eid = {L29},
        pages = {L29},
          doi = {10.3847/2041-8213/ace437},
archivePrefix = {arXiv},
       eprint = {2307.11406},
 primaryClass = {astro-ph.SR},
       adsurl = {https://ui.adsabs.harvard.edu/abs/2023ApJ...952L..29W}
}

@ARTICLE{Jiang-2007SoPh..240...77J,
       author = {{Jiang}, Yunchun and {Chen}, Huadong and {Shen}, Yuandeng and {Yang}, Liheng and {Li}, Kejun},
        title = "{H{\ensuremath{\alpha}} Dimming Associated With the Eruption of a Coronal Sigmoid in the Quiet Sun}",
      journal = {\solphys},
         year = 2007,
        month = jan,
       volume = {240},
       number = {1},
        pages = {77-87},
          doi = {10.1007/s11207-006-0257-4},
       adsurl = {https://ui.adsabs.harvard.edu/abs/2007SoPh..240...77J}
}

@ARTICLE{Titov-1999A&A...351..707T,
       author = {{Titov}, V.~S. and {D{\'e}moulin}, P.},
        title = "{Basic topology of twisted magnetic configurations in solar flares}",
      journal = {\aap},
         year = 1999,
        month = nov,
       volume = {351},
        pages = {707-720},
       adsurl = {https://ui.adsabs.harvard.edu/abs/1999A&A...351..707T}
}

@ARTICLE{Kannan-2024MNRAS.533.3611K,
       author = {{Kannan}, Arjun and {Yadav}, Nitin},
        title = "{Vortex dynamics in various solar magnetic field configurations}",
      journal = {\mnras},
         year = 2024,
        month = sep,
       volume = {533},
       number = {3},
        pages = {3611-3622},
          doi = {10.1093/mnras/stae1990},
archivePrefix = {arXiv},
       eprint = {2408.08225},
 primaryClass = {astro-ph.SR},
       adsurl = {https://ui.adsabs.harvard.edu/abs/2024MNRAS.533.3611K}
}

@ARTICLE{Xie-2025ApJ...979...27X,
       author = {{Xie}, Quan and {Liu}, Jiajia and {Nelson}, Chris J. and {Erd{\'e}lyi}, Robert and {Wang}, Yuming},
        title = "{Photospheric Swirls in a Quiet-Sun Region}",
      journal = {\apj},
         year = 2025,
        month = jan,
       volume = {979},
       number = {1},
          eid = {27},
        pages = {27},
          doi = {10.3847/1538-4357/ad99d4},
archivePrefix = {arXiv},
       eprint = {2412.03816},
 primaryClass = {astro-ph.SR},
       adsurl = {https://ui.adsabs.harvard.edu/abs/2025ApJ...979...27X}
}

@ARTICLE{Liu-2019ApJ...872...22L,
       author = {{Liu}, Jiajia and {Nelson}, Chris J. and {Erd{\'e}lyi}, Robertus},
        title = "{Automated Swirl Detection Algorithm (ASDA) and Its Application to Simulation and Observational Data}",
      journal = {\apj},
         year = 2019,
        month = feb,
       volume = {872},
       number = {1},
          eid = {22},
        pages = {22},
          doi = {10.3847/1538-4357/aabd34},
archivePrefix = {arXiv},
       eprint = {1804.02931},
 primaryClass = {astro-ph.SR},
       adsurl = {https://ui.adsabs.harvard.edu/abs/2019ApJ...872...22L}
}

@ARTICLE{Xie-2025A&A...700A...6X,
       author = {{Xie}, Quan and {Liu}, Jiajia and {Erd{\'e}lyi}, Robert and {Wang}, Yuming},
        title = "{Improving the {\ensuremath{\Gamma}}-functions method for vortex identification}",
      journal = {\aap},
         year = 2025,
        month = aug,
       volume = {700},
          eid = {A6},
        pages = {A6},
          doi = {10.1051/0004-6361/202554524},
archivePrefix = {arXiv},
       eprint = {2505.14384},
 primaryClass = {astro-ph.SR},
       adsurl = {https://ui.adsabs.harvard.edu/abs/2025A&A...700A...6X}
}

@ARTICLE{Xing-2024A&A...682A...3X,
       author = {{Xing}, Chen and {Aulanier}, Guillaume and {Schmieder}, Brigitte and {Cheng}, Xin and {Ding}, Mingde},
        title = "{Identifying footpoints of pre-eruptive and coronal mass ejection flux ropes with sunspot scars}",
      journal = {\aap},
         year = 2024,
        month = feb,
       volume = {682},
          eid = {A3},
        pages = {A3},
          doi = {10.1051/0004-6361/202347053},
archivePrefix = {arXiv},
       eprint = {2310.13532},
 primaryClass = {astro-ph.SR},
       adsurl = {https://ui.adsabs.harvard.edu/abs/2024A&A...682A...3X}
}

@ARTICLE{Robinson-2023A&A...673A..79R,
       author = {{Robinson}, Rebecca A. and {Aulanier}, Guillaume and {Carlsson}, Mats},
        title = "{Quiet Sun flux rope formation via incomplete Taylor relaxation}",
      journal = {\aap},
         year = 2023,
        month = may,
       volume = {673},
          eid = {A79},
        pages = {A79},
          doi = {10.1051/0004-6361/202346065},
archivePrefix = {arXiv},
       eprint = {2303.11738},
 primaryClass = {astro-ph.SR},
       adsurl = {https://ui.adsabs.harvard.edu/abs/2023A&A...673A..79R}
}

@ARTICLE{Robinson-2022A&A...668A.177R,
       author = {{Robinson}, Rebecca A. and {Carlsson}, Mats and {Aulanier}, Guillaume},
        title = "{From incoherent field to coherent reconnection. Understanding convection-driven coronal heating in the quiet Sun}",
      journal = {\aap},
         year = 2022,
        month = dec,
       volume = {668},
          eid = {A177},
        pages = {A177},
          doi = {10.1051/0004-6361/202244750},
archivePrefix = {arXiv},
       eprint = {2211.08364},
 primaryClass = {astro-ph.SR},
       adsurl = {https://ui.adsabs.harvard.edu/abs/2022A&A...668A.177R}
}

@ARTICLE{Amari-2025ApJ...984L..37A,
       author = {{Amari}, Tahar and {Canou}, Aur{\'e}lien and {Velli}, Marco and {Mikic}, Zoran and {Alauzet}, Frederic and {Buchlin}, Eric and {Luciani}, Jean-Fran{\c{c}}ois and {Aly}, Jean-Jacques and {Tarr}, Lucas A.},
        title = "{The Ubiquity of Twisted Flux Ropes in the Quiet Sun}",
      journal = {\apjl},
         year = 2025,
        month = may,
       volume = {984},
       number = {1},
          eid = {L37},
        pages = {L37},
          doi = {10.3847/2041-8213/adb74f},
archivePrefix = {arXiv},
       eprint = {2411.10563},
 primaryClass = {astro-ph.SR},
       adsurl = {https://ui.adsabs.harvard.edu/abs/2025ApJ...984L..37A}
}

@ARTICLE{Kuniyoshi-2023ApJ...949....8K,
       author = {{Kuniyoshi}, Hidetaka and {Shoda}, Munehito and {Iijima}, Haruhisa and {Yokoyama}, Takaaki},
        title = "{Magnetic Tornado Properties: A Substantial Contribution to the Solar Coronal Heating via Efficient Energy Transfer}",
      journal = {\apj},
         year = 2023,
        month = may,
       volume = {949},
       number = {1},
          eid = {8},
        pages = {8},
          doi = {10.3847/1538-4357/accbb8},
archivePrefix = {arXiv},
       eprint = {2304.03010},
 primaryClass = {astro-ph.SR},
       adsurl = {https://ui.adsabs.harvard.edu/abs/2023ApJ...949....8K}
}

@ARTICLE{Kuniyoshi-2024ApJ...969L..34K,
       author = {{Kuniyoshi}, Hidetaka and {Bose}, Souvik and {Yokoyama}, Takaaki},
        title = "{Comprehensive Synthesis of Magnetic Tornado: Cospatial Incidence of Chromospheric Swirls and Extreme-ultraviolet Brightening}",
      journal = {\apjl},
         year = 2024,
        month = jul,
       volume = {969},
       number = {2},
          eid = {L34},
        pages = {L34},
          doi = {10.3847/2041-8213/ad5a0e},
archivePrefix = {arXiv},
       eprint = {2404.18892},
 primaryClass = {astro-ph.SR},
       adsurl = {https://ui.adsabs.harvard.edu/abs/2024ApJ...969L..34K}
}

@ARTICLE{Kuniyoshi-2025ApJ...990L..71K,
       author = {{Kuniyoshi}, Hidetaka and {Imada}, Shinsuke and {Yokoyama}, Takaaki},
        title = "{A Unified Picture of Swirl-driven Solar Coronal Heating: Magnetic Energy Supply and Dissipation}",
      journal = {\apjl},
         year = 2025,
        month = sep,
       volume = {990},
       number = {2},
          eid = {L71},
        pages = {L71},
          doi = {10.3847/2041-8213/adfc61},
archivePrefix = {arXiv},
       eprint = {2508.14459},
 primaryClass = {astro-ph.SR},
       adsurl = {https://ui.adsabs.harvard.edu/abs/2025ApJ...990L..71K}
}

@ARTICLE{Skirvin-2025ApJ...988...18S,
       author = {{Skirvin}, Samuel J. and {Fedun}, Viktor and {Verth}, Gary and {Ballai}, Istvan},
        title = "{3D Magnetohydrodynamic Wave Propagation and Energy Transport in a Simulated Solar Vortex}",
      journal = {\apj},
         year = 2025,
        month = jul,
       volume = {988},
       number = {1},
          eid = {18},
        pages = {18},
          doi = {10.3847/1538-4357/adce73},
archivePrefix = {arXiv},
       eprint = {2504.12745},
 primaryClass = {astro-ph.SR},
       adsurl = {https://ui.adsabs.harvard.edu/abs/2025ApJ...988...18S}
}

@ARTICLE{Moreno-2025ApJ...995..137M,
       author = {{Moreno-Insertis}, Fernando and {Hansteen}, Viggo H. and {N{\'o}brega-Siverio}, Daniel},
        title = "{Magnetic Flux Cancellation in the Solar Atmosphere through 3D Realistic Numerical Modeling}",
      journal = {\apj},
         year = 2025,
        month = dec,
       volume = {995},
       number = {2},
          eid = {137},
        pages = {137},
          doi = {10.3847/1538-4357/ae14fb},
archivePrefix = {arXiv},
       eprint = {2510.19993},
 primaryClass = {astro-ph.SR},
       adsurl = {https://ui.adsabs.harvard.edu/abs/2025ApJ...995..137M}
}

@ARTICLE{Barra-2008AdSpR..42..917B,
       author = {{Barra}, Vincent and {Delouille}, V{\'e}ronique and {Hochedez}, Jean-Fran{\c{c}}ois},
        title = "{Segmentation of extreme ultraviolet solar images via multichannel fuzzy clustering}",
      journal = {Advances in Space Research},
         year = 2008,
        month = sep,
       volume = {42},
       number = {5},
        pages = {917-925},
          doi = {10.1016/j.asr.2007.10.021},
       adsurl = {https://ui.adsabs.harvard.edu/abs/2008AdSpR..42..917B}
}

@ARTICLE{Barra-2009A&A...505..361B,
       author = {{Barra}, V. and {Delouille}, V. and {Kretzschmar}, M. and {Hochedez}, J.-F.},
        title = "{Fast and robust segmentation of solar EUV images: algorithm and results for solar cycle 23}",
      journal = {\aap},
         year = 2009,
        month = oct,
       volume = {505},
       number = {1},
        pages = {361-371},
          doi = {10.1051/0004-6361/200811416},
       adsurl = {https://ui.adsabs.harvard.edu/abs/2009A&A...505..361B}
}

@ARTICLE{Verbeeck-2014A&A...561A..29V,
       author = {{Verbeeck}, C. and {Delouille}, V. and {Mampaey}, B. and {De Visscher}, R.},
        title = "{The SPoCA-suite: Software for extraction, characterization, and tracking of active regions and coronal holes on EUV images}",
      journal = {\aap},
         year = 2014,
        month = jan,
       volume = {561},
          eid = {A29},
        pages = {A29},
          doi = {10.1051/0004-6361/201321243},
       adsurl = {https://ui.adsabs.harvard.edu/abs/2014A&A...561A..29V}
}

@ARTICLE{Inceoglu-2022ApJ...930..118I,
       author = {{Inceoglu}, Fadil and {Shprits}, Yuri Y. and {Heinemann}, Stephan G. and {Bianco}, Stefano},
        title = "{Identification of Coronal Holes on AIA/SDO Images Using Unsupervised Machine Learning}",
      journal = {\apj},
         year = 2022,
        month = may,
       volume = {930},
       number = {2},
          eid = {118},
        pages = {118},
          doi = {10.3847/1538-4357/ac5f43},
archivePrefix = {arXiv},
       eprint = {2203.10491},
 primaryClass = {astro-ph.SR},
       adsurl = {https://ui.adsabs.harvard.edu/abs/2022ApJ...930..118I}
}

@ARTICLE{Panos-2018ApJ...861...62P,
       author = {{Panos}, Brandon and {Kleint}, Lucia and {Huwyler}, Cedric and {Krucker}, S{\"a}m and {Melchior}, Martin and {Ullmann}, Denis and {Voloshynovskiy}, Sviatoslav},
        title = "{Identifying Typical Mg II Flare Spectra Using Machine Learning}",
      journal = {\apj},
         year = 2018,
        month = jul,
       volume = {861},
       number = {1},
          eid = {62},
        pages = {62},
          doi = {10.3847/1538-4357/aac779},
archivePrefix = {arXiv},
       eprint = {1805.10494},
 primaryClass = {astro-ph.SR},
       adsurl = {https://ui.adsabs.harvard.edu/abs/2018ApJ...861...62P}
}

@ARTICLE{Viticchie-2011A&A...530A..14V,
       author = {{Viticchi{\'e}}, B. and {S{\'a}nchez Almeida}, J.},
        title = "{Asymmetries of the Stokes V profiles observed by HINODE SOT/SP in the quiet Sun}",
      journal = {\aap},
         year = 2011,
        month = jun,
       volume = {530},
          eid = {A14},
        pages = {A14},
          doi = {10.1051/0004-6361/201016096},
archivePrefix = {arXiv},
       eprint = {1103.1987},
 primaryClass = {astro-ph.SR},
       adsurl = {https://ui.adsabs.harvard.edu/abs/2011A&A...530A..14V}
}

@ARTICLE{Amari-2015Natur.522..188A,
       author = {{Amari}, Tahar and {Luciani}, Jean-Fran{\c{c}}ois and {Aly}, Jean-Jacques},
        title = "{Small-scale dynamo magnetism as the driver for heating the solar atmosphere}",
      journal = {\nat},
         year = 2015,
        month = jun,
       volume = {522},
       number = {7555},
        pages = {188-191},
          doi = {10.1038/nature14478},
       adsurl = {https://ui.adsabs.harvard.edu/abs/2015Natur.522..188A}
}

@ARTICLE{ChenYJ-2021A&A...656L...7C,
       author = {{Chen}, Yajie and {Przybylski}, Damien and {Peter}, Hardi and {Tian}, Hui and {Auch{\`e}re}, F. and {Berghmans}, D.},
        title = "{Transient small-scale brightenings in the quiet solar corona: A model for campfires observed with Solar Orbiter}",
      journal = {\aap},
         year = 2021,
        month = dec,
       volume = {656},
          eid = {L7},
        pages = {L7},
          doi = {10.1051/0004-6361/202140638},
archivePrefix = {arXiv},
       eprint = {2104.10940},
 primaryClass = {astro-ph.SR},
       adsurl = {https://ui.adsabs.harvard.edu/abs/2021A&A...656L...7C}
}

@ARTICLE{Mou-2018A&A...619A..55M,
       author = {{Mou}, Chauzhou and {Madjarska}, Maria S. and {Galsgaard}, Klaus and {Xia}, Lidong},
        title = "{Eruptions from quiet Sun coronal bright points. I. Observations}",
      journal = {\aap},
         year = 2018,
        month = nov,
       volume = {619},
          eid = {A55},
        pages = {A55},
          doi = {10.1051/0004-6361/201833243},
archivePrefix = {arXiv},
       eprint = {1808.04541},
 primaryClass = {astro-ph.SR},
       adsurl = {https://ui.adsabs.harvard.edu/abs/2018A&A...619A..55M}
}

@article{Rempel-2017,
	adsurl = {https://ui.adsabs.harvard.edu/abs/2017ApJ...834...10R},
	archiveprefix = {arXiv},
	author = {{Rempel}, M.},
	doi = {10.3847/1538-4357/834/1/10},
	eid = {10},
	eprint = {1609.09818},
	journal = {\apj},
	month = {Jan},
	number = {1},
	pages = {10},
	primaryclass = {astro-ph.SR},
	title = {{Extension of the MURaM Radiative MHD Code for Coronal Simulations}},
	volume = {834},
	year = {2017}}

@ARTICLE{BakSteslicka-2013ApJ...770L..28B,
       author = {{B{\k{a}}k-St{\k{e}}{\'s}licka}, Urszula and {Gibson}, Sarah E. and {Fan}, Yuhong and {Bethge}, Christian and {Forland}, Blake and {Rachmeler}, Laurel A.},
        title = "{The Magnetic Structure of Solar Prominence Cavities: New Observational Signature Revealed by Coronal Magnetometry}",
      journal = {\apjl},
         year = 2013,
        month = jun,
       volume = {770},
       number = {2},
          eid = {L28},
        pages = {L28},
          doi = {10.1088/2041-8205/770/2/L28},
archivePrefix = {arXiv},
       eprint = {1304.7388},
 primaryClass = {astro-ph.SR},
       adsurl = {https://ui.adsabs.harvard.edu/abs/2013ApJ...770L..28B}
}

@ARTICLE{Ruminska-2022ApJ...926..146R,
       author = {{Rumi{\'n}ska}, Agnieszka and {B{\k{a}}k-St{\c{e}}{\'s}licka}, Urszula and {Gibson}, Sarah E. and {Fan}, Yuhong},
        title = "{Coronal Cavities in CoMP Observations}",
      journal = {\apj},
         year = 2022,
        month = feb,
       volume = {926},
       number = {2},
          eid = {146},
        pages = {146},
          doi = {10.3847/1538-4357/ac469c},
       adsurl = {https://ui.adsabs.harvard.edu/abs/2022ApJ...926..146R}
}

@ARTICLE{Gou-2024SoPh..299...99G,
       author = {{Gou}, Tingyu and {Liu}, Rui and {Su}, Yang and {Veronig}, Astrid M. and {Pan}, Hanya and {Luo}, Runbin and {Gan}, Weiqun},
        title = "{High-Resolution Observation of Blowout Jets Regulated by Sunspot Rotation}",
      journal = {\solphys},
         year = 2024,
        month = jul,
       volume = {299},
       number = {7},
          eid = {99},
        pages = {99},
          doi = {10.1007/s11207-024-02333-8},
archivePrefix = {arXiv},
       eprint = {2406.02783},
 primaryClass = {astro-ph.SR},
       adsurl = {https://ui.adsabs.harvard.edu/abs/2024SoPh..299...99G}
}

@ARTICLE{Kitiashvili-2012ApJ...751L..21K,
       author = {{Kitiashvili}, I.~N. and {Kosovichev}, A.~G. and {Mansour}, N.~N. and {Wray}, A.~A.},
        title = "{Dynamics of Magnetized Vortex Tubes in the Solar Chromosphere}",
      journal = {\apjl},
         year = 2012,
        month = may,
       volume = {751},
       number = {1},
          eid = {L21},
        pages = {L21},
          doi = {10.1088/2041-8205/751/1/L21},
archivePrefix = {arXiv},
       eprint = {1201.5442},
 primaryClass = {astro-ph.SR},
       adsurl = {https://ui.adsabs.harvard.edu/abs/2012ApJ...751L..21K}
}

@ARTICLE{Kitiashvili-2013ApJ...770...37K,
       author = {{Kitiashvili}, I.~N. and {Kosovichev}, A.~G. and {Lele}, S.~K. and {Mansour}, N.~N. and {Wray}, A.~A.},
        title = "{Ubiquitous Solar Eruptions Driven by Magnetized Vortex Tubes}",
      journal = {\apj},
         year = 2013,
        month = jun,
       volume = {770},
       number = {1},
          eid = {37},
        pages = {37},
          doi = {10.1088/0004-637X/770/1/37},
archivePrefix = {arXiv},
       eprint = {1301.0018},
 primaryClass = {astro-ph.SR},
       adsurl = {https://ui.adsabs.harvard.edu/abs/2013ApJ...770...37K}
}
\bibliographystyle{aasjournalv7}
\end{document}